\documentclass[traditabstract]{aa}

\pdfoutput=1
\usepackage{graphicx}
\usepackage[font=small]{caption}
\usepackage{indentfirst}
\usepackage{tabu}
\usepackage[percent]{overpic}
\usepackage{transparent}
\usepackage{ragged2e}

\usepackage{natbib}
\usepackage{booktabs}
\usepackage{fixltx2e}

\usepackage{amsmath, amssymb}
\usepackage{txfonts}
\usepackage[a]{esvect}
\usepackage{tikz}

\newcommand{\degree}[0]{$^{\circ}$}

\newcommand{\cbra}[1]{\left( #1 \right)}      % put the argument between parentheses (curve brackets)
\newcommand{\sbra}[1]{\left[ #1 \right]}      % put the argument between square brackets
\newcommand{\avg}[1]{\left< #1 \right>}         % average
\let\baraccent=\=
\renewcommand{\=}[1]{\stackrel{#1}{=}} % for putting numbers above =
\let\arrowaccent=\>
\renewcommand{\>}[1]{\stackrel{#1}{\Rightarrow}} % for putting numbers above =>

\makeatletter
\newcommand{\rmnum}[1]{{\footnotesize{\expandafter\@slowromancap\romannumeral #1@}}}
\newcommand{\Rmnum}[1]{{\expandafter\@slowromancap\romannumeral #1@}}
\makeatother

\everymath={\rm}

\begin{document}

\title{Atomic-to-molecular gas phase transition triggered by the radio jet in Centaurus A\thanks{This publication is based on data acquired with the Atacama Pathfinder Experiment (APEX) under programme ID 096.B-0892.}}

\author{
   Q. Salom\'e\inst{1}  \and
   P. Salom\'e\inst{1}  \and
   F. Combes\inst{1,2}  \and
   S. Hamer\inst{3}
}

\institute{
   LERMA, Observatoire de Paris, CNRS, UPMC, PSL Univ., 61 avenue de l'Observatoire, 75014 Paris, France \\ email: quentin.salome@obspm.fr \and
   Coll\`ege de France, 11 place Marcelin Berthelot, 75005 Paris \and
   CRAL, Observatoire de Lyon, CNRS, Universit\'e Lyon 1, 9 Avenue Ch. Andr\'e, 69561 Saint Genis Laval cedex, France
}

\date{Received ??? / Accepted ??}

\titlerunning{APEX data of the outer filaments of Centaurus A}
\authorrunning{Salomé et al.}

\abstract{
   NGC 5128 (Centaurus A) is one of the best example to study AGN-feedback in the local Universe. At 13.5 kpc from the galaxy, optical filaments with recent star formation are lying along the radio-jet direction. We used the Atacama Pathfinder EXperiment (APEX) to map the CO(2-1) emission all along the filaments structure. Molecular gas mass of $(8.2\pm 0.5)\times 10^7\: M_\odot$ was found over the 4.2 kpc-structure which represents about 3\% of the total gas mass of the NGC 5128 cold gas content. Two dusty mostly molecular structures are identified, following the optical filaments. The region corresponds to the crossing of the radio jet with the northern H\rmnum{1} shell, coming from a past galaxy merger. One filament is located at the border of the H\rmnum{1} shell, while the other is entirely molecular, and devoid of H\rmnum{1} gas. The molecular mass is comparable to the H\rmnum{1} mass in the shell, suggesting a scenario where the atomic gas was shocked and transformed in molecular clouds by the radio jet.
Comparison with combined FIR Herschel and UV GALEX estimation of star formation rates in the same regions leads to depletion times of more than 10 Gyr. The filaments are thus less efficient than discs in converting molecular gas into stars. Kinetic energy injection triggered by shocks all along the jet/gas interface is a possible process that appears to be consistent with MUSE line ratio diagnostics derived in a smaller region of the northern filaments. Whether the AGN is the sole origin of this energy input and what is the dominant (mechanical vs radiative) mode for this process is however still to be investigated.}


\maketitle

%%%%%%%%%%%%%%%%%%%%%%%%%%%%%%%%%%%%%%%%%%%%%%%%%%%%%%%%%%%%%%%%%%%%%%%%%%%%%%%%%%%%%%%%%%%%%%%%%%%%%%%%%%%%

\section{Introduction}

   Understanding the detailed processes of the radio jets-ISM/ICM interaction is a key missing piece in the scenario of AGN-regulated galaxy growth. In radio galaxies, interaction between radio jets and the surrounding ISM is suspected to regulate star formation (\textit{negative} feedback; \citealt{Bower_2006, Croton_2006}). But it has also sometimes been claimed to locally enhance the star formation (\textit{positive} feedback; \citealt{Croft_2006,Bogdan_2011}).

   Interaction of a jet with molecular gas is very likely present along filaments surrounding NGC 5128 (also known as Centaurus A). This giant nearby early type galaxy lies at the heart of a moderately rich group of galaxies. It hosts a massive disc of dust, gas and young stars in its central regions \citep{Israel_1998}. This disc of gas presents a misalignment in CO \citep{Espada_2009}, likely due to a recent merger event. NGC 5128 is surrounded by faint arc-like stellar shells (at a radius of several kpc around the galaxy) where H\rmnum{1} gas has been detected \citep{Schiminovich_1994} and CO emission has been observed at the intersection with the radio jet \citep{Charmandaris_2000}. In addition, \cite{Auld_2012} detected large amount of dust ($\sim 10^5\: M_\odot$) around the northern shell region.

   Optically bright filaments are observed in the direction of the radio jet \citep{Blanco_1975,Graham_1981,Morganti_1991}. These filaments are thought to be the place of star formation as confirmed by GALEX data \citep{Auld_2012} and young stars \citep{Rejkuba_2001}. These so-called inner and outer filaments are located at a distance of $\sim 7.7\: kpc$ and $\sim 13.5\: kpc$, respectively. The inner filament lies at the top of the inner radio lobe and could be the result of a weak cocoon-driven bow shock that propagates through the diffuse interstellar medium, triggering star formation \citep{Crockett_2012}. Huge X-ray filaments are present in the northern middle radio lobe \citep{Kraft_2009}, at the north of the outer filaments, and could  result from a jet-cloud interaction where cold, dense clouds have been shock heated to X-ray temperatures. Finally, the inner and outer filaments show distinct kinematical components, a well-defined knotty filament and a more diffuse structure, as highlighted by optical excitation lines (VIMOS and MUSE; \citealt{Santoro_2015a,Santoro_2015b,Hamer_2015}). Recently \cite{Santoro_2016} identified a star forming cloud in the MUSE data that contains several H\rmnum{2} regions. Some of these regions are currently forming stars whereas star formation seems to have recently stopped in the others.

\begin{figure*}[h!]
  \centering
  \includegraphics[height=11cm,trim=270 80 550 160,clip=true]{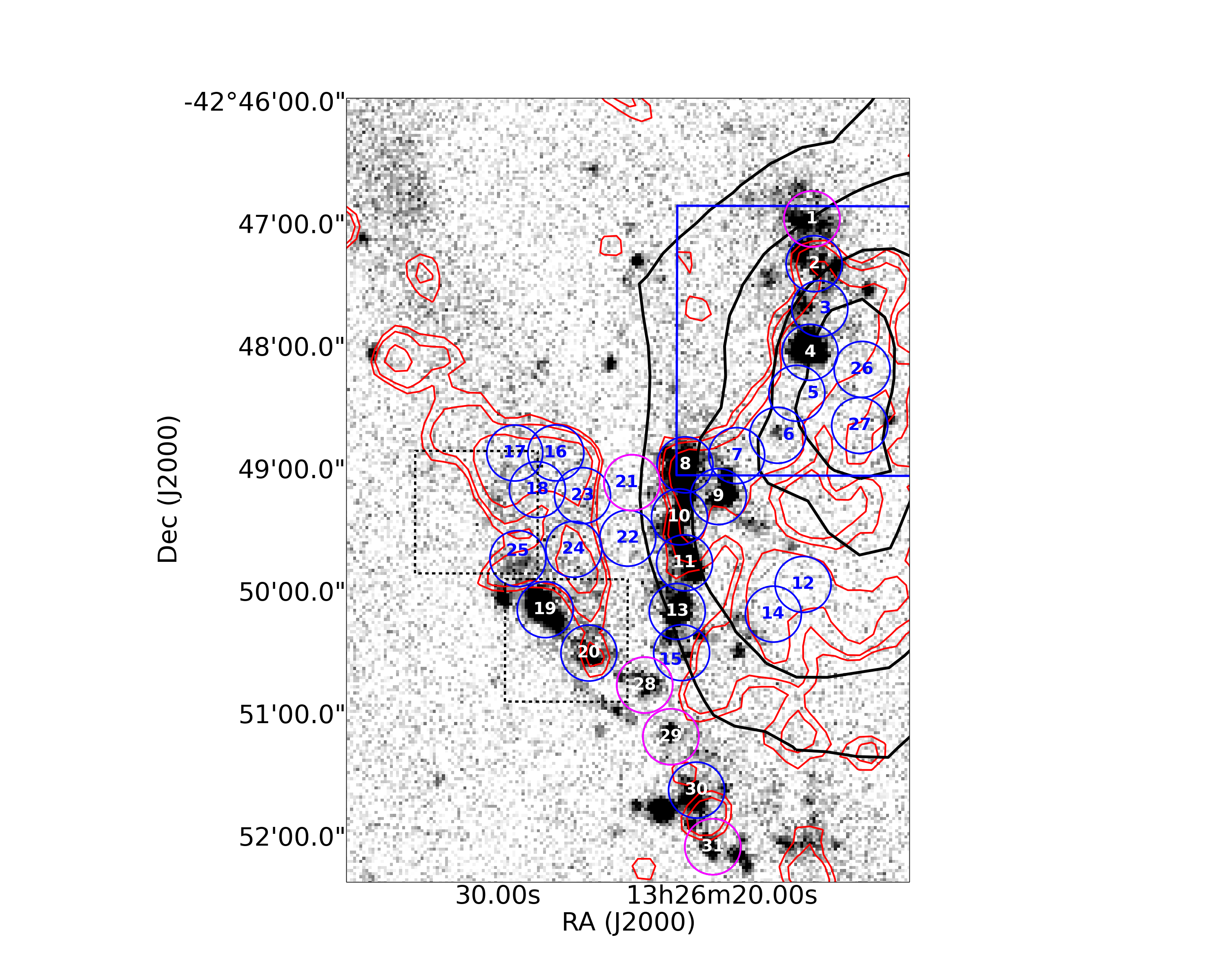}
  \hspace{5mm}
  \includegraphics[height=11cm,trim=237 30 210 62,clip=true]{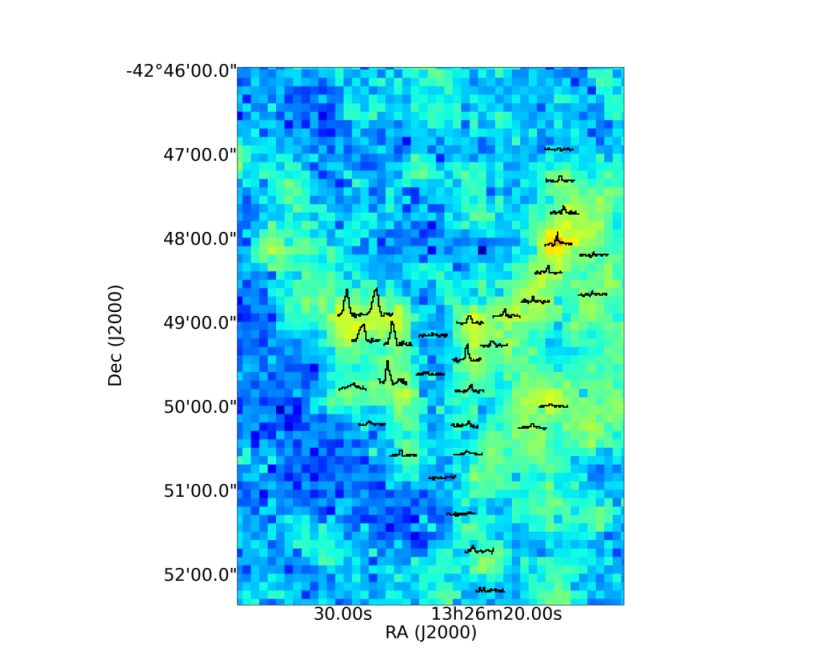}
  \caption{\label{overview} \emph{Left:} FUV image of the outer filaments from GALEX \citep{Neff_2015b}. The black and red contours correspond to the H\rmnum{1} and the Herschel-SPIRE $250\: \mu m$ emission \citep{Schiminovich_1994,Auld_2012}, respectively. The blue box corresponds the region observed by \cite{Charmandaris_2000} with SEST, and the dashed boxes show the field of view of MUSE observations \citep{Santoro_2015b}. The circles show the position observed with APEX (CO(2-1) beam). The magenta circles shows the non-detections. The regions of Table \ref{table:specCO} are labelled by their number. \emph{Right:} Map of the Herschel-SPIRE $250\: \mu m$ emission. The APEX spectra are overlaid in black.}
\end{figure*}

   In \cite{SalomeQ_2016}, we conducted a study of the outer filaments based on archival data. We compared the molecular gas reservoir (via CO with SEST and ALMA archival data) to star formation tracers (FUV with GALEX, FIR with Herschel). We concentrated on a small region of $130''\sim 2\: kpc$ and found that star formation seems to be very inefficient, with large depletion times. In addition, ALMA data revealed the presence of 3 distinct unresolved and dynamically separated clumps. 
The virial parameter $\alpha_{vir}=5\sigma_c^2\, R_c/(GM_c)$ \citep{Bertoldi_1992} of the ALMA clumps ($\sim 10-16$) indicates that an input of kinetic energy may have occurred and could explain why these clouds appear inefficient to form stars despite their surface density $N_{H_2}\geq 10^{20}\: cm^{-2}$.
Here we extend this study to larger scales ($\sim 255''\sim 4.2\: kpc$), based on new CO(2-1) observations with APEX.

   For consistency with our previous paper on Centaurus A \citep{SalomeQ_2016}, we used the distance of 3.42 Mpc derived by \cite{Ferrarese_2007}, leading to a scaling conversion of $16.5\: pc/''$. However recent review of measurements by \cite{Harris_2010} led to a more precise value of 3.8 Mpc. This difference translates into a 10\% difference in the spatial scale, and an underestimate of about 20\% of the masses and star formation rates. Nevertheless it does not change the star formation efficiency.

\section{Observations}
\label{sec:Obs}

   Millimetre observations of the CO(2-1) emission were made with the APEX telescope in September 2015 (20 positions) and December 2015 (11 positions). At redshift z=0.001826, this line is observable at a frequency of 230.118 GHz, which leads to a primary beam of $27.4''\sim 450\: pc$. We mapped the whole filaments with 31 pointings. The observations were made with the SHeFI/APEX-1 receiver\footnote{http://www.apex-telescope.org/heterodyne/shfi/het230/} and backends XFFTS (bandwidths of 2.5 GHz; resolution of 88.5 kHz). The typical system temperature was $145-180\: K$.

   The data were reduced using the IRAM package CLASS. After dropping bad spectra, a linear baseline was subtracted from the average spectrum; for detections, the baseline was subtracted at velocities outside the range of the emission line. Then, each spectrum was smoothed to a spectral resolution of $\sim 12.5\: km/s$.
During a first run (September), 20 positions were observed with a homogeneous rms of $\sim 3\: mK$. Some of the positions were then re-observed (second run) to improve the sensitivity to a rms of $\sim 2\: mK$. Finally, 11 new positions were observed during the second run with a rms of $\sim 2-3\: mK$ except if the signal was strong enough (positions 23 and 24).
The observations are summarised in Table \ref{table:obs} and the resulting spectra are plotted in Fig. \ref{spectra}.

\section{Results}
\label{sec:Res}

   \subsection{CO emission}

\begin{figure*}[h]
  \centering
  \includegraphics[height=9cm,trim=340 100 420 155,clip=true]{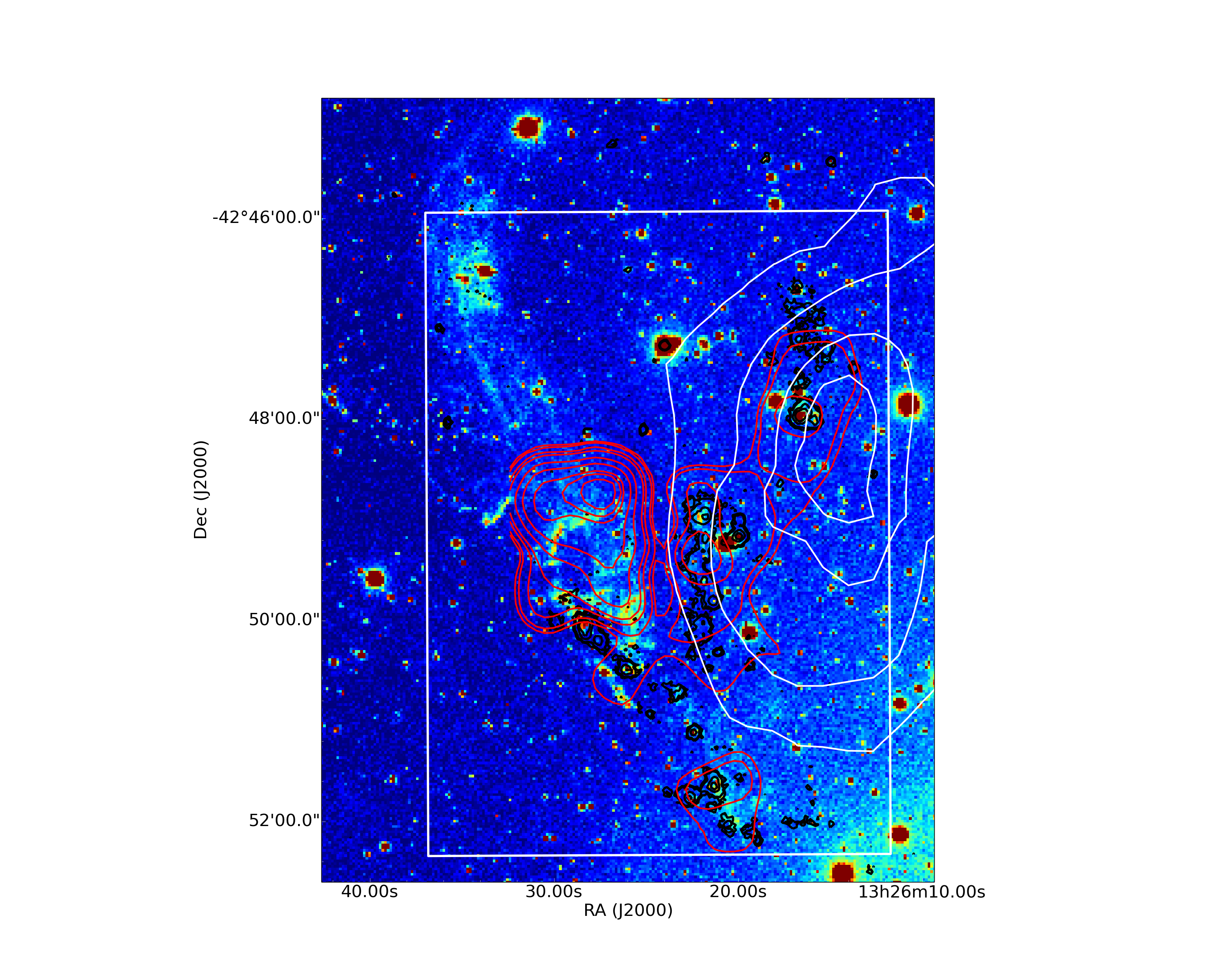}
  \includegraphics[height=9cm,trim=200 80 455 155,clip=true]{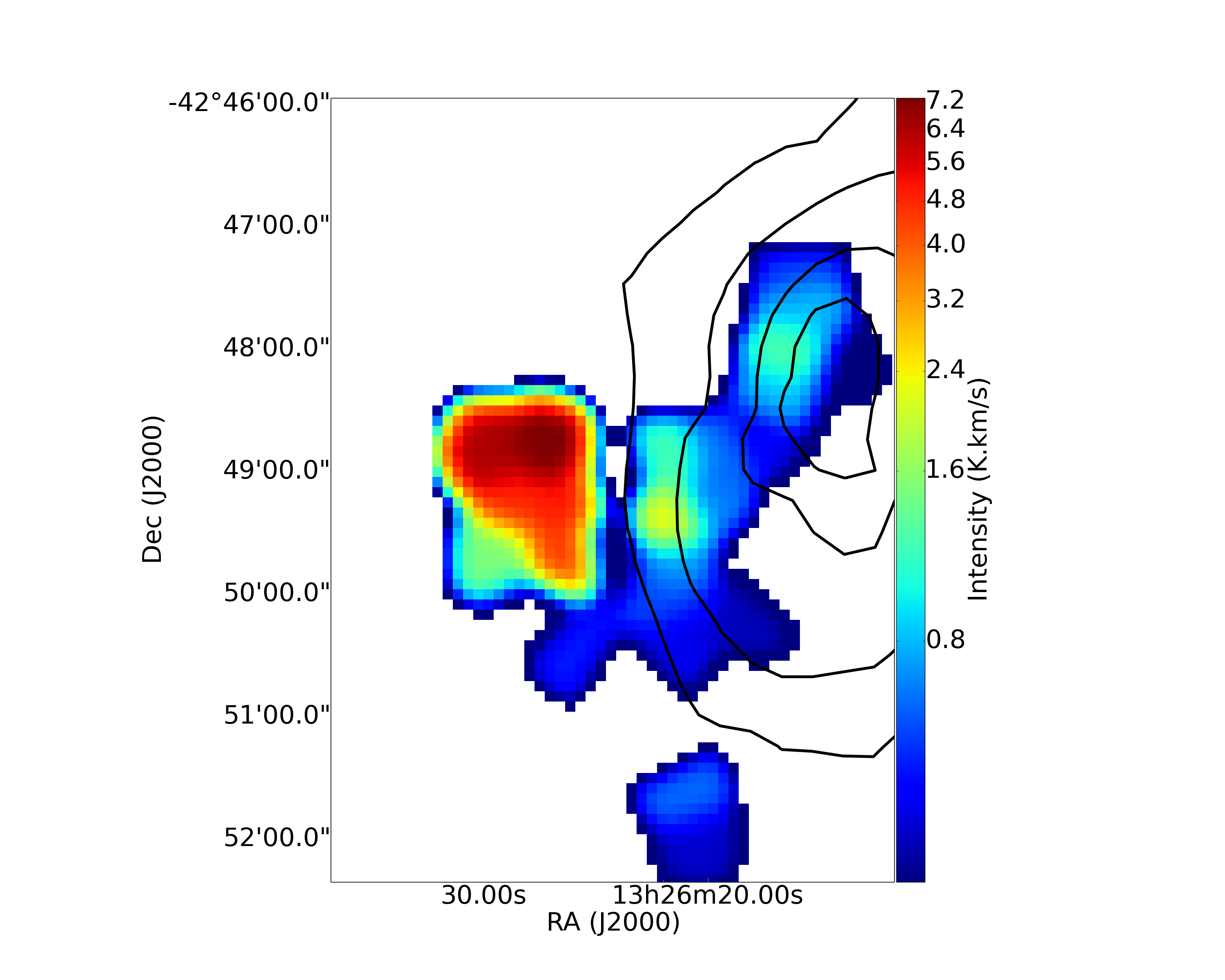}
  \caption{\label{Ha_APEX} \emph{Left:} $H\alpha$ emission of the northern region of Centaurus A with CTIO. We overlaid the H\rmnum{1} emission (VLA; white contours), the CO(2-1) emission (APEX; red contours) and the FUV emission (GALEX; black contours). \emph{Right:} Intensity map of the CO(2-1) emission from APEX in $K\,km\,s^{-1}$, with the H\rmnum{1} emission overlaid in black contours. The white box on the left shows the region of the right panel.}
\end{figure*}

      \subsubsection{CO detection and star formation}

   CO emission was detected for 26/31 positions. The CO detections follow the dust emission and are distributed along the UV filaments.
The undetected positions are regions 1, 21, 28, 29, 31 (see Fig \ref{overview} and Table \ref{table:specCO}) and all lie outside the dusty area revealed by Herschel. Region 1 is at the top-north, close to the ALMA detected clumps, in a region where the CO emission is fading, at the edge of the dusty area. Region 21 is between two main East vs. West CO-bright regions. This suggests that those two main regions are clearly separated. However the nearby region 22 is well-detected in CO. The two branches of the northern filaments seen in optical may thus hide an underlying continuous molecular reservoir. More sensitive observations will confirm this possibility. Regions 28, 29, 31 constitute a continuous North-South region at the very bottom tip of the filaments. The southern part of the filaments seems to be CO free or at least CO poor. One region (30) is however detected in this structure. It corresponds to the position of the only bright star forming UV-clump of the southern structure. \\
Some positions detected in CO do not show UV emission, probably due to dust extinction. In addition, positions 16-20 and 23-25 also present $H\alpha$ emission (CTIO; see Fig. \ref{Ha_APEX}). But the lack of $H\alpha$ emission in the rest of the filaments can be explained by the narrow band filter of CTIO, the emission being outside the observing band.
CO emission is stronger in the eastern part of the filaments (right panel of Figure \ref{overview}). The larger molecular mass ($9-12\times 10^6\: M_\odot$) that we derived in this eastern region indicates that several GMC associations are located there.

   Estimating the $L'_{CO}$ with the formula from \cite{Solomon_1997} and the CO(2-1)/CO(1-0) ratio of 0.55 from \cite{Charmandaris_2000}, and applying a standard Milky Way $\alpha_{CO}=4.6\: M_\odot\,(K\,km\,s^{-1}\,pc^2)^{-1}$ \citep{Solomon_1997}, we derived molecular gas masses of a few $10^5-10^7\: M_\odot$. For non detections, we calculated an upper limit at $3\sigma$ with a line width of $30\: km\,s^{-1}$, similar to the neighbour positions. \\
The star formation rate (SFR) has been derived from the FUV (GALEX) and IR (Herschel) emission in regions of $27.4''$ too (see \cite{SalomeQ_2016} for the details). We also derived the molecular depletion times $t_{dep}^{mol}=M_{H_2}/SFR$ of all APEX positions. Detailed results for each position are summarised in Tables \ref{table:specCO} and \ref{table:overview}.
In the whole region that has been observed with APEX, we found a total molecular gas mass $M_{H_2}=(8.2\pm 0.5)\times 10^7\: M_\odot$, about five times higher that the mass found by \cite{Charmandaris_2000}. Although the region presented here is larger than that of \cite{Charmandaris_2000}, we did not expect such difference in mass as they observed the peak of H\rmnum{1} emission. The difference comes from the large amount of gas found outside the H\rmnum{1} cloud.
A combination of the IR and FUV emission gives a star formation rate $SFR=(1.08\pm 0.12)\times 10^{-3}\: M_\odot\,yr^{-1}$. This gives a molecular depletion time in the filaments $t_{dep}=75.5\pm 13.0\: Gyr$, indicating that the star formation is very inefficient in the filaments.

      \subsubsection{Comparison between ALMA and APEX data}

   Position 1 has the same coordinates as the ALMA data published by \cite{SalomeQ_2016}. However, the three detected clumps lie outside the primary beam of ALMA. They are therefore contained in position 2. The ALMA emission was claimed to be $S_{CO}\Delta v\sim 3.0\: Jy\,km\,s^{-1}$ and the position 2 with APEX has an intensity $I_{CO}\sim 0.636\: K\,km\,s^{-1}$. The APEX Jy/K conversion factor of 32.5 gives a CO flux $S_{CO}\Delta v\sim 20.7\: Jy\,km\,s^{-1}$ thus the ratio of the APEX and ALMA fluxes is $\sim 7$.
If we assume that the CO emission only consists of clumps then there could be $\sim 7$ times more clumps in the APEX beam. Therefore position 2 would contain $\sim 21$ clumps of mass $\sim 5\times 10^4\: M_\odot$\footnote{There is a mistake in \cite{SalomeQ_2016}. We underestimated the mass of the clumps as we forgot to take into account the CO(2-1)/CO(1-0) ratio.}.
The ALMA data can resolve spatial scales up to $\sim 18''$ (baselines of $18.4-250.8\: m$) therefore the signal is unlikely filtered by the small baseline coverage.

      \subsubsection{Molecular-to-atomic mass ratio}

   The outer filaments lie in projection at the interface between the radio jet of Centaurus A and the H\rmnum{1} shell. Using the VLA data from \cite{Schiminovich_1994}, we could derive $H_2$/H\rmnum{1} mass ratios at different positions in the filaments. As the resolution of the VLA data ($40''\times 78''$) is lower than the APEX data, we combined several pointings from APEX that are contained in a single VLA beam. The eastern part of the filaments was not detected in H\rmnum{1} therefore we used the rms of 2 mJy and a line width of $80\: km\,s^{-1}$ \citep{Schiminovich_1994} to derive upper limits at $3\sigma$. We conclude that the filaments are mostly molecular (see Table \ref{table:H2-HI}), except for positions 12-14 and 26-27 that lie outside the UV emitting region.

\begin{table}[h]
  \centering
  \footnotesize
  \begin{tabular}{lcccc}
    \hline \hline
    Positions   & H\rmnum{1} flux  &  $M_{H\rmnum{1}}$ &     $M_{H_2}$     & $M_{H_2}/M_{H\rmnum{1}}$ \\
                & ($Jy.km.s^{-1}$) &    ($M_\odot$)    &    ($M_\odot$)    &                          \\ \hline
    1-2-3       &       1.01       &  $2.8\times 10^6$ & $<2.8\times 10^6$ &         $<1.00$          \\
    4-5-6       &       1.68       &  $4.6\times 10^6$ &  $4.6\times 10^6$ &           1.00           \\
    7-8-9-10    &       0.62       &  $1.7\times 10^6$ &  $9.4\times 10^6$ &           5.42           \\
    12-14       &       0.94       &  $2.6\times 10^6$ &  $1.8\times 10^6$ &           0.69           \\
    11-13-15    &       0.30       &  $8.2\times 10^5$ &  $4.5\times 10^6$ &           5.49           \\
    26-27       &       2.01       &  $5.5\times 10^6$ &  $1.1\times 10^6$ &           0.20           \\
    29-30       &       0.21       &  $5.9\times 10^5$ & $<1.8\times 10^6$ &         $<3.04$          \\ \hline
    West        &       6.77       &  $1.9\times 10^7$ &  $2.6\times 10^7$ &       $\sim 1.37$        \\ \hline
    16-17-18    &     $<0.33$      & $<9.2\times 10^5$ &  $3.2\times 10^7$ &         $>35.1$          \\
    19-20-24    &     $<0.34$      & $<9.3\times 10^5$ &  $9.0\times 10^6$ &         $>9.73$          \\
    21-22-23    &       0.20       &  $5.5\times 10^5$ & $<1.0\times 10^7$ &         $<18.2$          \\ \hline
    East        &       0.87       &  $2.4\times 10^6$ &  $5.1\times 10^7$ &       $\sim 21.3$        \\ \hline
    Full region &       7.64       &  $2.1\times 10^7$ &  $7.7\times 10^7$ &           3.66           \\ \hline
  \end{tabular}
  \caption{\label{table:H2-HI} H\rmnum{1} and $H_2$ masses in combinations of APEX pointings. The H\rmnum{1} fluxes are in $Jy\,km\,s^{-1}$, masses are in $M_\odot$.}
\end{table}

      \subsubsection{Probability distribution function}

\begin{figure}[h]
  \centering
  \includegraphics[width=\linewidth,trim=25 5 50 35,clip=true]{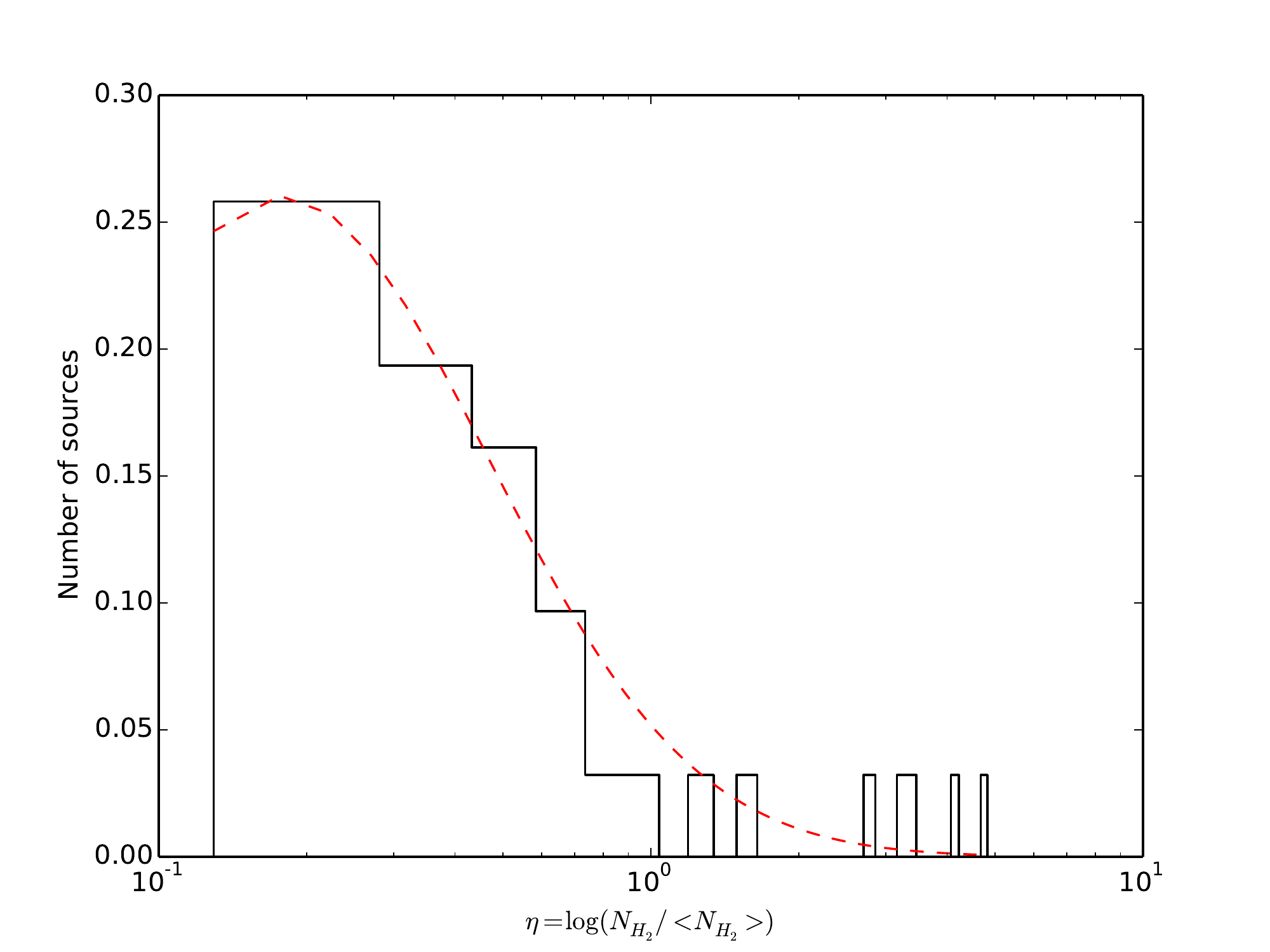}
  \caption{\label{PDF} Probability distribution function of the $H_2$ column density of the APEX pointings. The x-axis is the normalised column density. The red dashed line is the best fitting log-normal function (equation \ref{eq:PDF}) with $\sigma=0.96$ and $\mu=-0.80$.}
\end{figure}

   We derived the probability distribution function (PDF) of the $H_2$ column density of the APEX pointings (figure \ref{PDF}), following \cite{Druard_2014}. For the 31 APEX pointings, we derived the column density by assuming an homogeneous emission over $450\times 450\: pc^2$. Even if there is very few statistics and a lack of resolution, the PDF starts to follow a log-normal distribution:
\begin{equation} \label{eq:PDF}
  p_\eta d\eta=\frac{1}{\sqrt{2\pi \sigma^2}} \exp\sbra{-\frac{(\eta-\mu)^2}{2\sigma^2}}
\end{equation}

\noindent The best fitting result characterises the log-normal by the mean $\mu=-0.80$ and standard deviation $\sigma=0.96$. The log-normal shape of the PDF indicates that the structure of the molecular gas is dominated by turbulence. However, we do not have enough statistics at small scales to determine the contribution of gravitation (power-law).
This will be possible by resolving the GMC at high resolution with forthcoming ALMA observations.

   \subsection{An inefficient star formation}

\begin{figure}[h]
\centering
  \begin{overpic}[page=2,width=\linewidth,trim=25 0 45 35,clip=true]{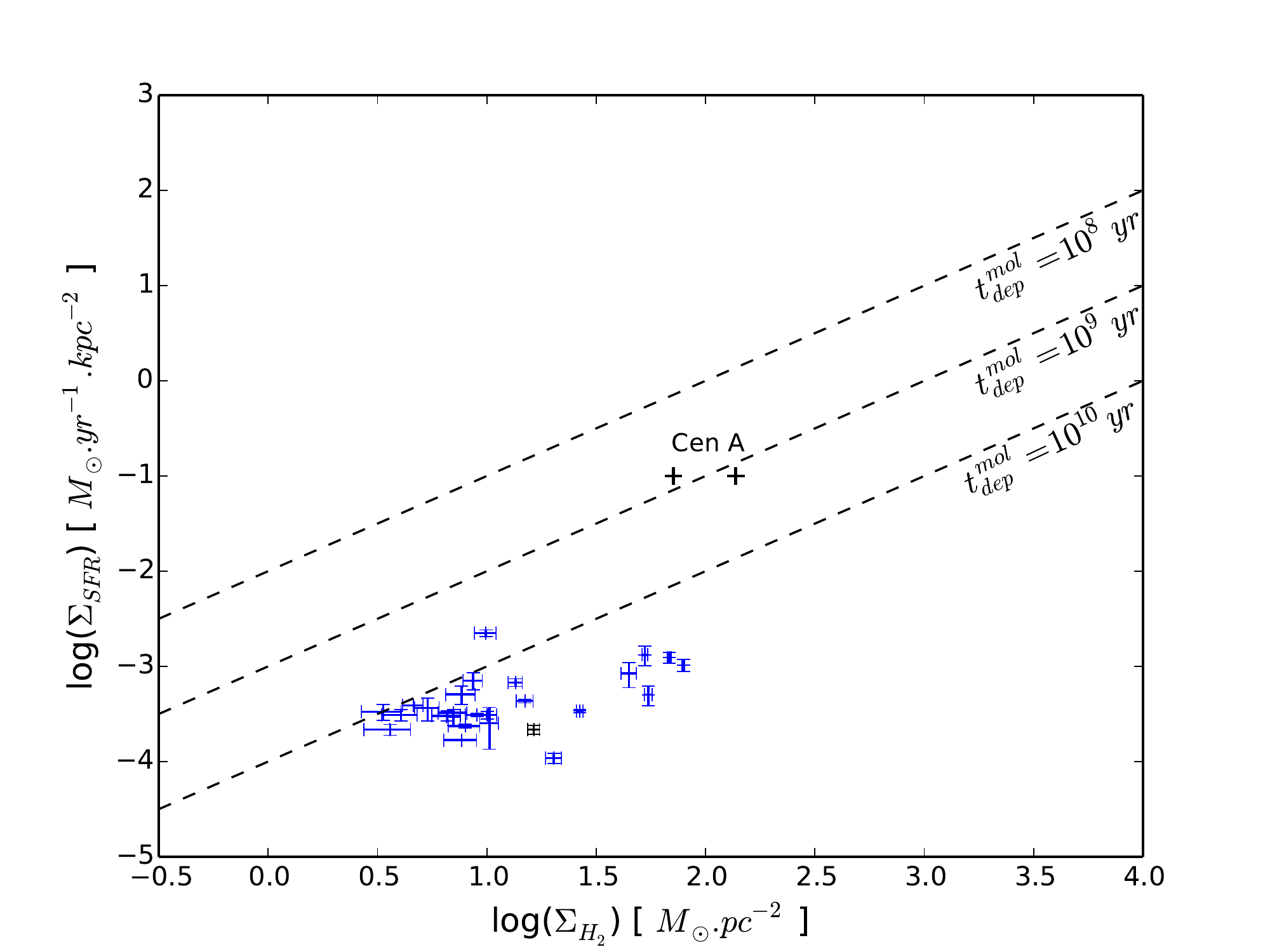}
    \put(32.5,25.7){\transparent{0.5}\includegraphics[width=26mm,height=13.7mm]{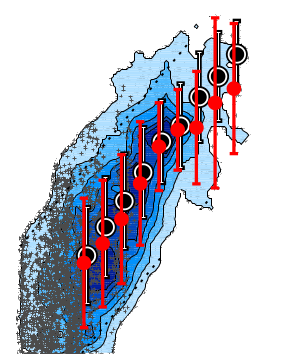}}
  \end{overpic}
  \caption{\label{KS-law} $\Sigma_{SFR}$ vs. $\Sigma_{H_2}$ for the different regions of CO emission observed with APEX (blue). The black crosses correspond to the central galaxy and the entire filaments ($\Sigma_{H_2}\sim 16.4\: M_\odot\,pc^{-2}$; $\Sigma_{SFR}\sim 2.17\times 10^{-4}\: M_\odot\,yr^{-1}\,kpc^{-2}$). The diagonal dashed lines show lines of constant molecular gas depletion times of, from top to bottom, $10^8$, $10^9$, and $10^{10}\: yr$. We overlay the contours of \cite{Leroy_2013} for nearby spiral galaxies. The red and blue lines represent Equation \ref{eq:Renaud} for the different set of parameters presented below.}
\end{figure}

   We calculated the molecular gas and SFR surface densities ($\Sigma_{H_2}$, $\Sigma_{SFR}$) for all the positions (see Table \ref{table:overview}). Both quantities were smoothed over the APEX $27.4''$ beam. We then plotted the $\Sigma_{SFR}$ vs $\Sigma_{H_2}$ diagram (Figure \ref{KS-law}).
As mentioned before, the star formation seems very inefficient, with depletion times of the order of tens of Gyr. Nevertheless, the positions seem to follow a Schmidt-Kennicutt law $\Sigma_{SFR}\propto \Sigma_{H_2}^N$ \citep{Kennicutt_1998a}, lying lower than star forming spiral galaxies. \\
The outer filaments of Centaurus A contain large amounts of molecular gas with a small SFE. Thus star formation is rather inefficient in the filaments.
However, star formation tracers like dust and UV emission are definitely detected and only in the filaments. The presence of young stars \citep{Rejkuba_2001} also indicates that these regions are the place of recent star formation, triggered in a fairly large molecular gas reservoir.

   \cite{Renaud_2012} developed a model to study the effect of turbulence on the star formation efficiency. We adapted their model by hypothesising that the SFR volume density $\rho_{SFR}$ is simply proportional to the molecular gas volume density: $\rho_{SFR}=\epsilon_{SFE}\, \rho_{H_2}$ above a threshold gas density $\rho_0$. Without stellar feedback, the theoretical law between $\Sigma_{SFR}$ and $\Sigma_{H_2}$ for a gas distribution given by equation \ref{eq:PDF} is:
\begin{equation} \label{eq:Renaud}
  \Sigma_{SFR}=\frac{\epsilon_{SFE}\, \Sigma_{H_2}}{2} erfc\cbra{\frac{\ln \frac{\rho_0 h}{\Sigma_{H_2}}-\mu}{\sigma\sqrt{2}}}
\end{equation}

\begin{figure*}[h!]
  \centering
  \includegraphics[height=9cm,trim=200 80 407 160,clip=true]{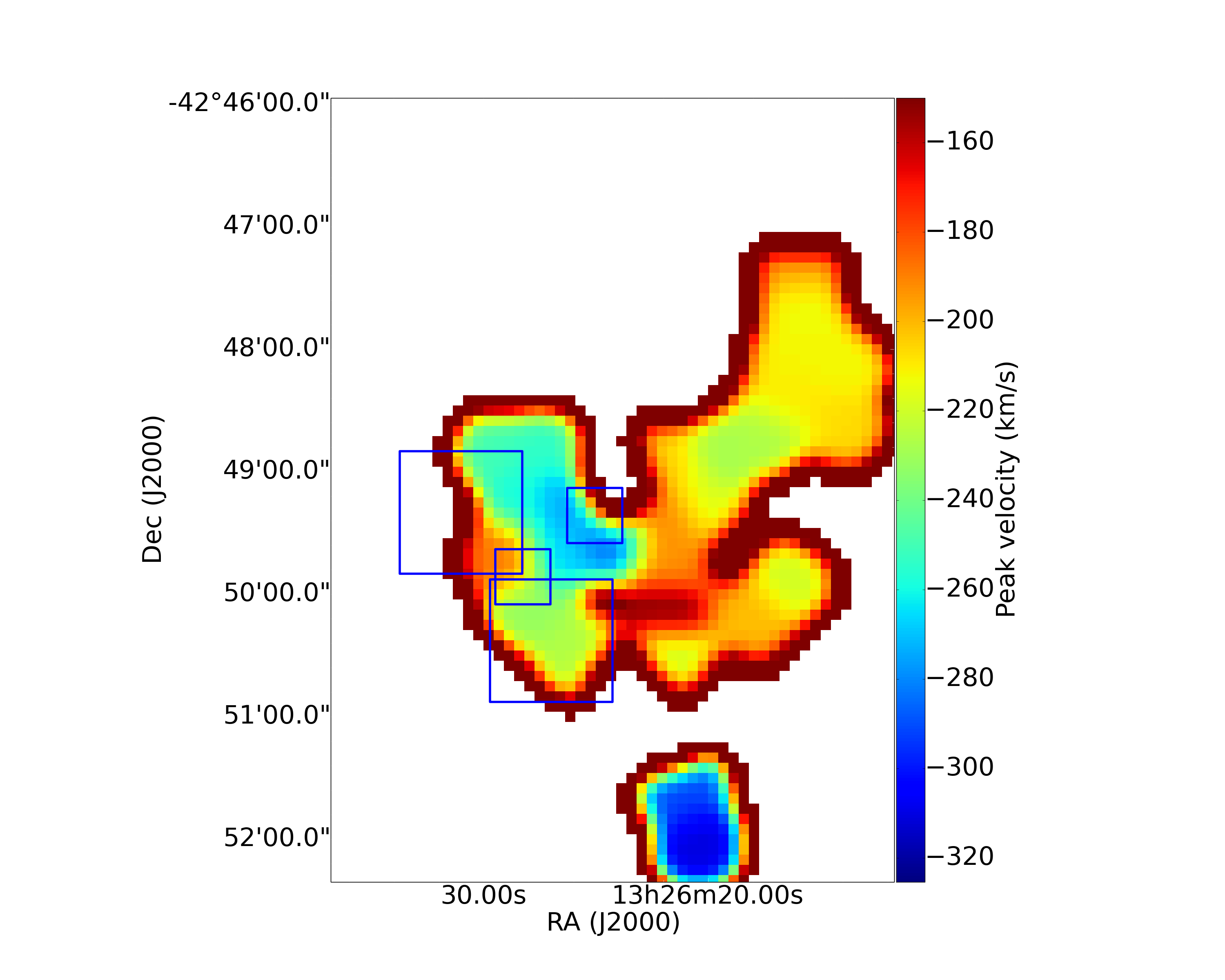}
  \hspace{3mm}
  \includegraphics[height=9cm,trim=580 80 445 160,clip=true]{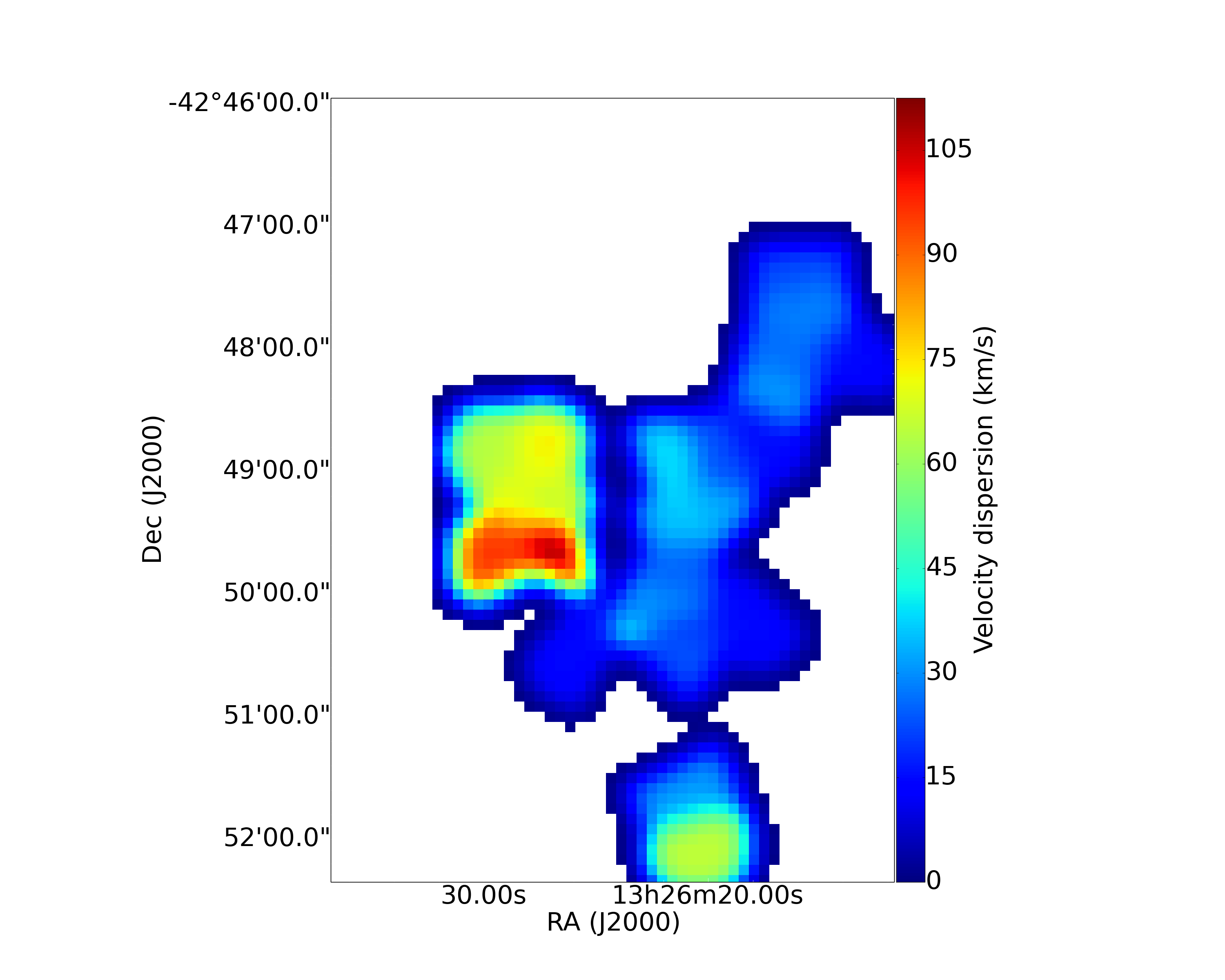}
  \caption{\label{CO-moments} Peak velocity (\emph{left}) and dispersion velocity (\emph{right}) map of the CO(2-1) emission from APEX. The velocities are relative to the systemic $547\: km\,s^{-1}$ value. In the left panel, the boxes represent the fields of view observed with VIMOS (small; \citealt{Santoro_2015a}) and MUSE (large; \citealt{Santoro_2015b}).}
\end{figure*}

\begin{figure*}[h!]
  \centering
  \includegraphics[height=8cm,trim=150 75 345 160,clip=true]{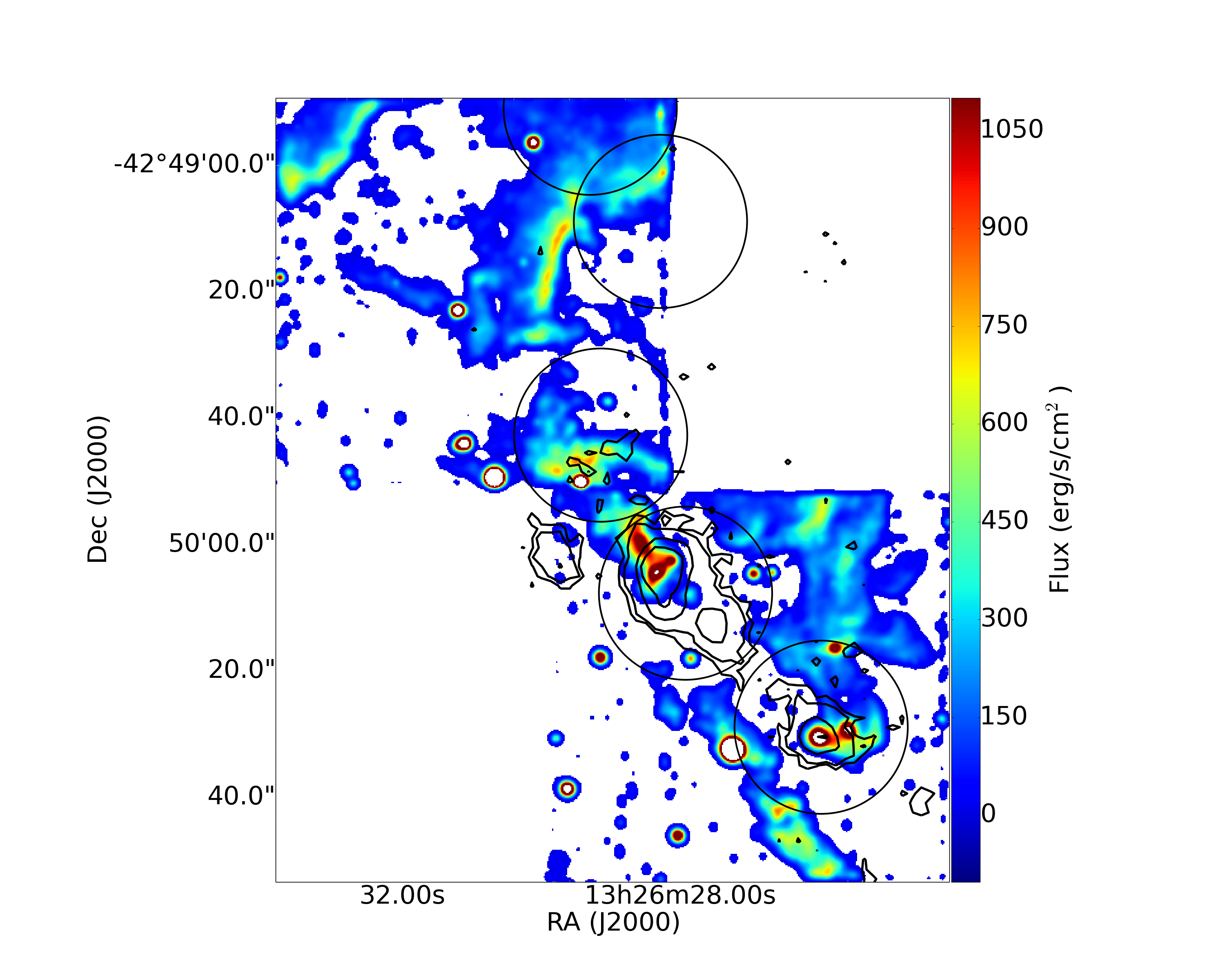}
  \hspace{5mm}
  \includegraphics[height=8cm,trim=485 75 310 160,clip=true]{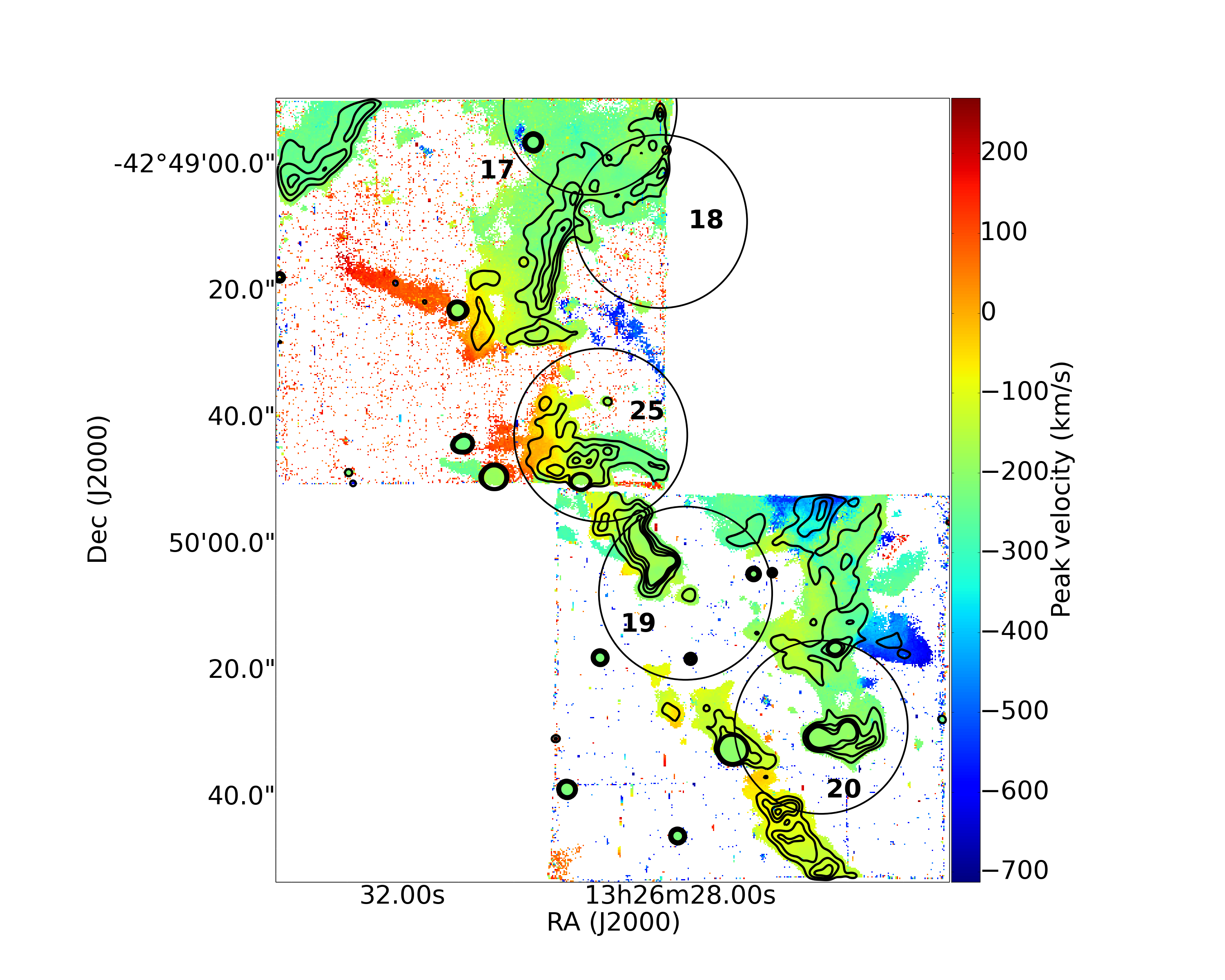}
  \caption{\label{MUSE-moments} \emph{Left:} $H\alpha$-[N\rmnum{2}] flux map, with the GALEX FUV emission overlaid in black contours. \emph{Right:} Peak velocity map of the $H\alpha$-[N\rmnum{2}] emission from MUSE, relative to Centaurus A. The contours shows the $H\alpha$-[N\rmnum{2}] flux. The field of view of the MUSE data corresponds to the larger blue boxes in Fig. \ref{CO-moments}. The APEX $27.4''$ beams are represented by the circles.}
\end{figure*}

\noindent We fixed $\epsilon_{SFE}=1/2\times 10^9\: yr^{-1}$ and $h=450\: pc$ (the size of the APEX beam), and explore different set of parameters ($\sigma$, $\mu$, $\rho_0$). We found that the region with the larger depletion times (the massive regions in the east) are best fitted by $\sigma=1.5$, $\mu=0.58$, $\rho_0=100\: cm^{-3}$, whereas the region in the west may be fitted by $\sigma=2.5$, $\mu=4.58$, $\rho_0=10^3\: cm^{-3}$. However the fits are highly degenerate and different set of parameters may fit the same regions (e.g. the blue dashed line in Fig \ref{KS-law}). This analysis needs better constrains on the PDF that will then constrain the parameters of equation \ref{eq:Renaud}.

   \subsection{Dynamics and morphology of the filaments}

   The overall CO(2-1) emission of the filaments is blueshifted compared to the central galaxy with peak velocities between $-200$ and $-300\: km\,s^{-1}$ in Centaurus A rest frame ($v_{hel}^{Cen\, A}=547\pm 5 \: km\,s^{-1}$; \citealt{Graham_1978}). The filaments show velocity gradients (figure \ref{CO-moments}), where the northern part of the filaments has the highest blueshifted velocities with respect to the systemic velocity. The eastern part presents higher velocities ($\avg{v}=-244.8\: km\,s^{-1}$) than the western region ($\avg{v}=-205.7\: km\,s^{-1}$). The CO(2-1) emission is also broader in the east with $\avg{\Delta v}=69.7\: km\,s^{-1}$, whereas the average velocity dispersion in the west is $\avg{\Delta v}=42.5\: km\,s^{-1}$. It is difficult to distinguish whether the velocity dispersion is produced by turbulence or by the proper motion of several clouds along the line of sight. The larger molecular mass (a few $10^6\: M_\odot$) combined with the broad emission lines in the east of the filaments indicates that the second solution is more probable. Incoming ALMA data will provide a clear view about the clump distribution.

   The APEX pointings do not accurately cover the MUSE observations. Nevertheless we observed part of the $H\alpha$ emission (Fig. \ref{MUSE-moments}). The right panel of Fig. \ref{MUSE-moments} shows the velocity of the $H\alpha$-[N\rmnum{2}] emission, relative to Centaurus A. At the location of the CO emission, part of the $H\alpha$ emission presents velocities of the order of the CO emission ones. The molecular and ionised components may thus be spatially and dynamically associated. A more accurate correlation requires a more homogeneous coverage of the CO emission at similar spatial resolution as the MUSE data.

\begin{figure}[h!]
  \centering
  \includegraphics[width=\linewidth,trim=220 10 490 90,clip=true]{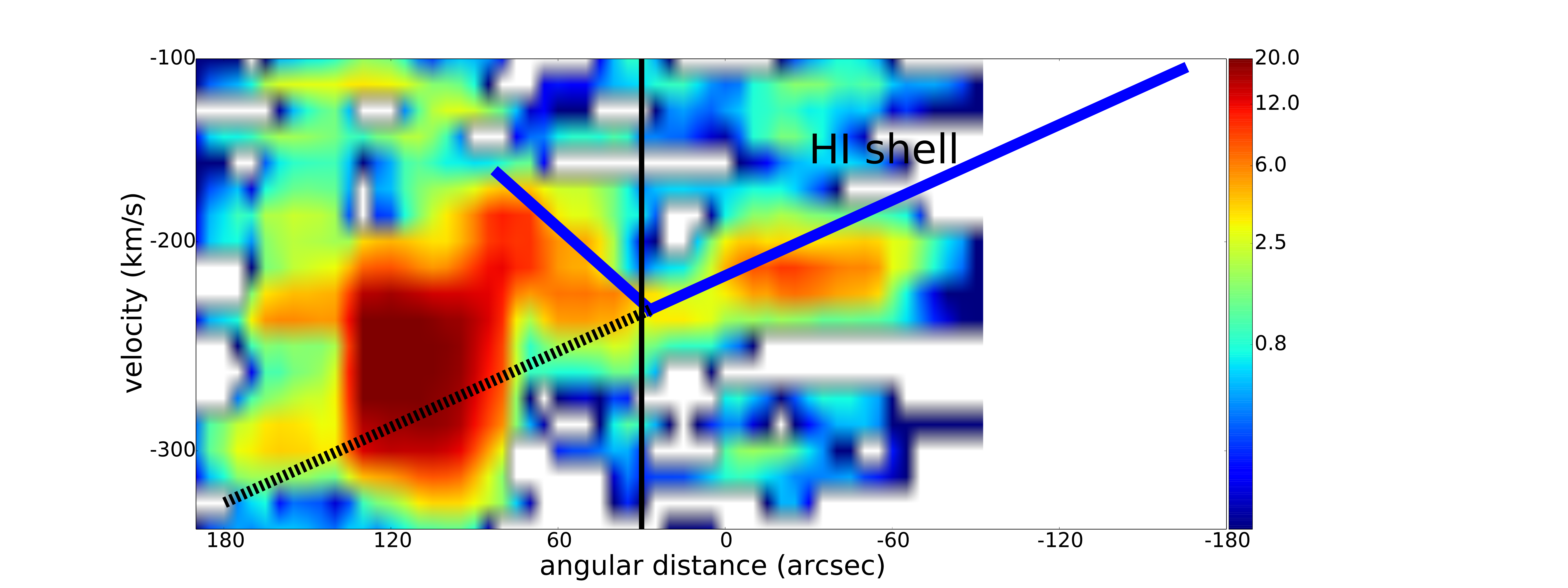}
  \caption{\label{gradient} PV diagram of the CO emission (in mK) centred in $\alpha=13^h 26^m 15^s$, $\delta=-$42:49:00 over the same slit orientation as \cite{Oosterloo_2005} with a width of $4.2'$ (taking all the CO emission). The blue lines represent the H\rmnum{1} cloud velocity gradient. The dashed line represents the continuity of this velocity gradient over the CO emission. The position of the radio jet is shown by the vertical black line.}
\end{figure}

   \cite{Oosterloo_2005} computed a PV diagram of the H\rmnum{1} gas along a slit oriented perpendicularly to the jet. The H\rmnum{1} cloud shows a velocity gradient all along its extension. At a distance of $30''-60''$ from the centre of the slit, the authors found a change in the slope of the velocity gradient that they interpreted as a direct effect of the jet interaction with the H\rmnum{1}. The CO data along the same slit (Fig. \ref{gradient}) show that (1) the large scale H\rmnum{1} velocity gradient is also seen in CO and that it seems to extend further out to the east side as shown by the bright CO spot at $\sim 120''$ (following the dashed line). (2) The break in the slope of the H\rmnum{1} velocity gradient is also seen in CO. Molecular and atomic gas thus seems to be dynamically associated.

   If, as claimed by \cite{Oosterloo_2005} and \cite{Santoro_2015a}, the change in velocity is due to the interaction of the radio-jet with the gas, then the gas projected velocity deceleration also affects the molecular components. The jet is expected to accelerate the gas but, a simple change in the orientation of the gas velocity vector induced by the jet can explain this behaviour, the higher velocity having a smaller projection along the line of sight.

   \subsection{Excitation of the filaments}

   The optically bright part of the outer filaments was observed with the Multi Unit Spectroscopic Explorer (MUSE) on the VLT \citep{MUSE} during the Sience Verification period (Program 60.A-9341(A) on 25 June 2014, \citealt{Santoro_2015b}). The observations consisted of 2 fields, each consisting of 3 pointings (6 pointings in total) of 1000s each, with a  90\degree rotation between each. We took the data from the archive and reduced them with version 0.18.1 of the MUSE data reduction pipeline and the European Southern Observatory Recipe Execution Tool (ESOREX v. 3.10.2) command-line interface. The final data cube was then sky subtracted using a $20''\times 20''$ region of the FOV free from line emission and stars to produce the sky model.
We extracted a cube that covers a velocity range of $120000\: km\,s^{-1}$. This cube covers all of the principal line complexes with a constant spectral sampling of $0.6\: \AA$ ($\sim 30\: km\,s^{-1}$ at the wavelength of $H\alpha$). We also extracted an individual cube that covers the $H\alpha$-[N\rmnum{2}] lines in the velocity range $-700$ to $250\: km\,s^{-1}$, relative to the systemic $547\: km\,s^{-1}$ value. This velocity range covers the three components identified by \cite{Santoro_2015b}, here in Fig. \ref{MUSE-moments}.
We corrected for pointing offsets after the rotation by matching the positions of bright stars within the field of view before combining the pointings from each field. The two fields were then combined to produce a single data cube.

\begin{figure*}[h!]
\centering
  \includegraphics[width=0.8\linewidth,trim=70 355 175 290,clip=true]{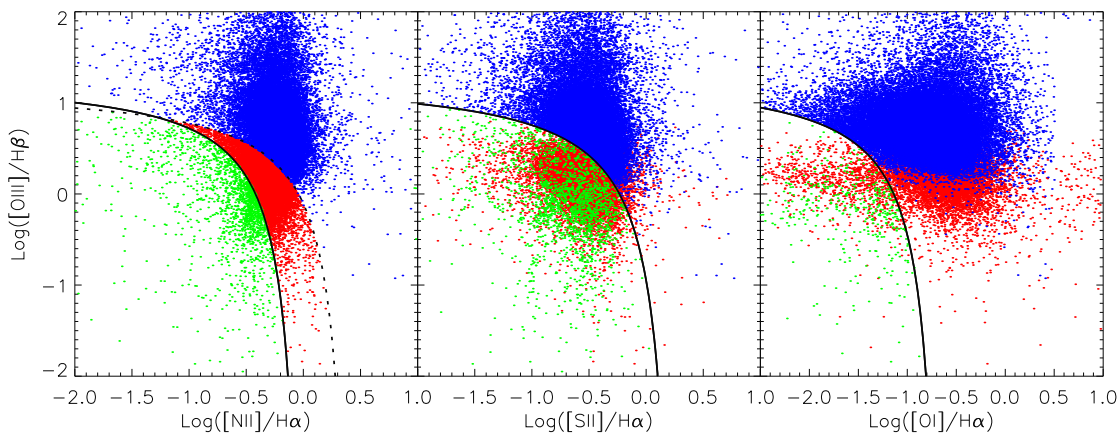}
  \caption{\label{BPT} Pixel-by-pixel BPT diagrams of the filaments with MUSE. The black line in the left panel represents the empirical separation of star formation (green) and AGN/shock-ionised regions (blue). The dotted line on the left and the solid lines in the other panels show the extreme upper limit for star formation \citep{Kewley_2006}. The red points correspond to the composite regime.}
\end{figure*}

\begin{figure*}[h!]
\centering
  \includegraphics[height=7cm,trim=160 80 450 170,clip=true]{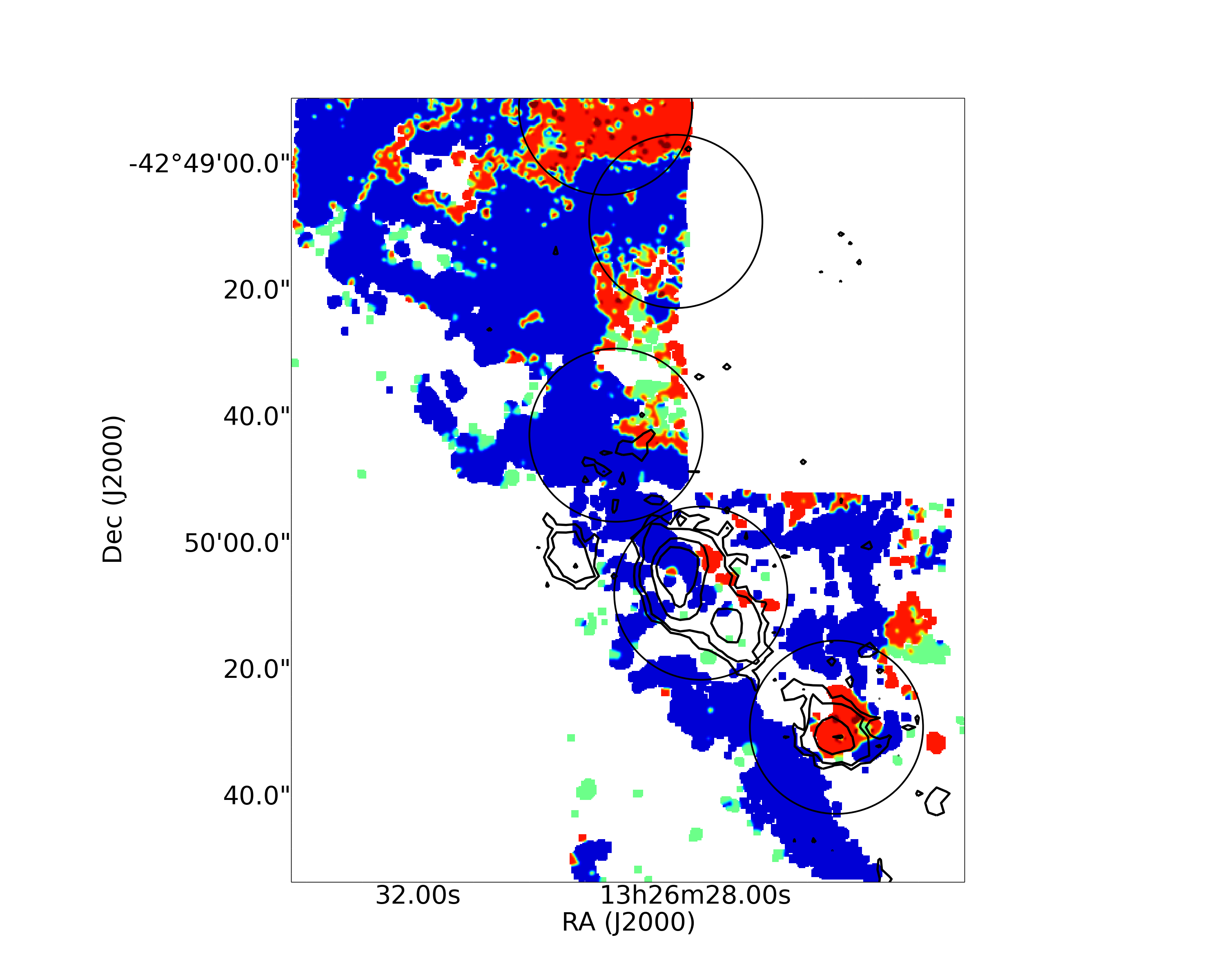}
  \hspace{5mm}
  \includegraphics[height=7cm,trim=510 80 450 170,clip=true]{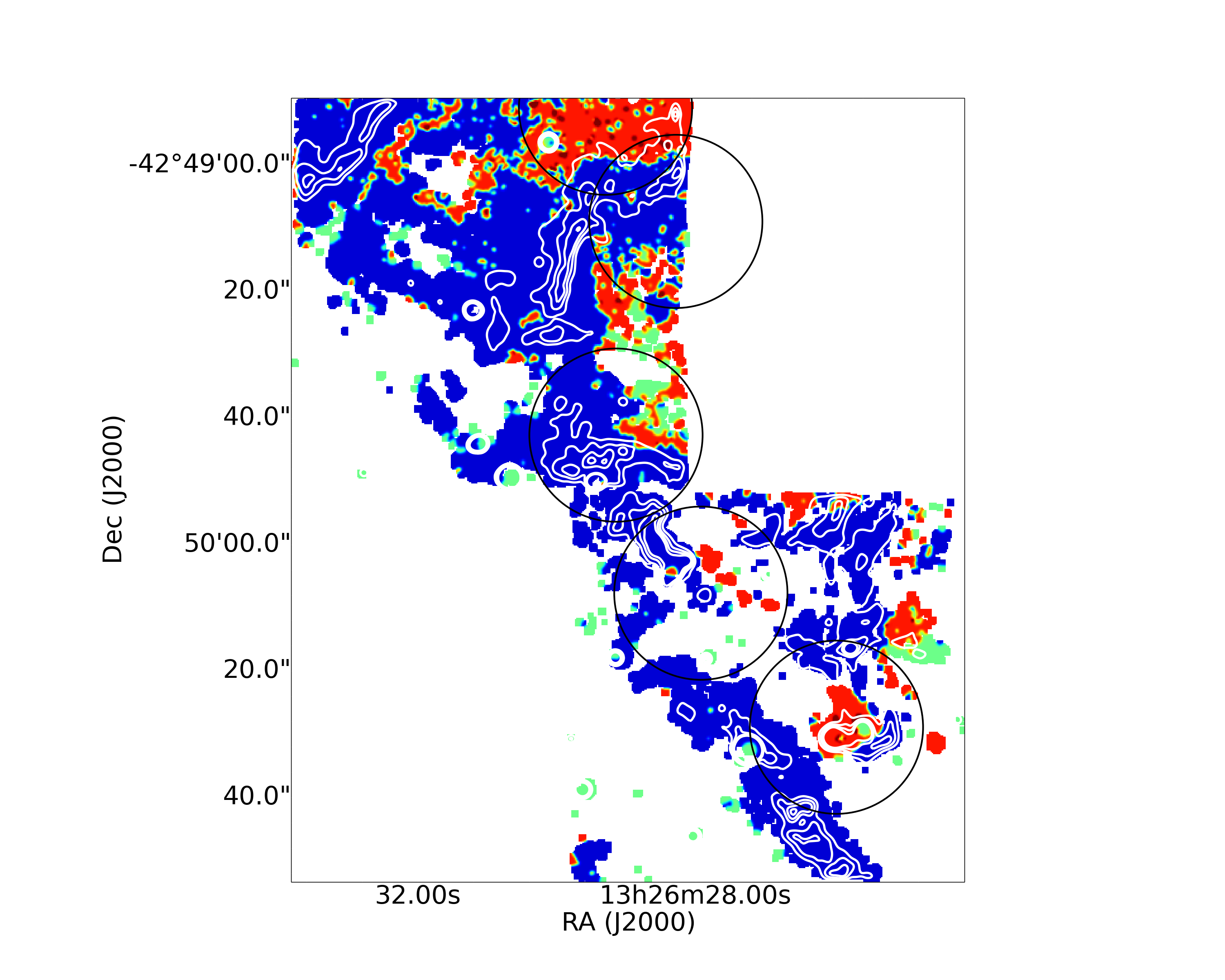}
  \vspace{5mm}
  \includegraphics[height=7cm,trim=160 80 450 170,clip=true]{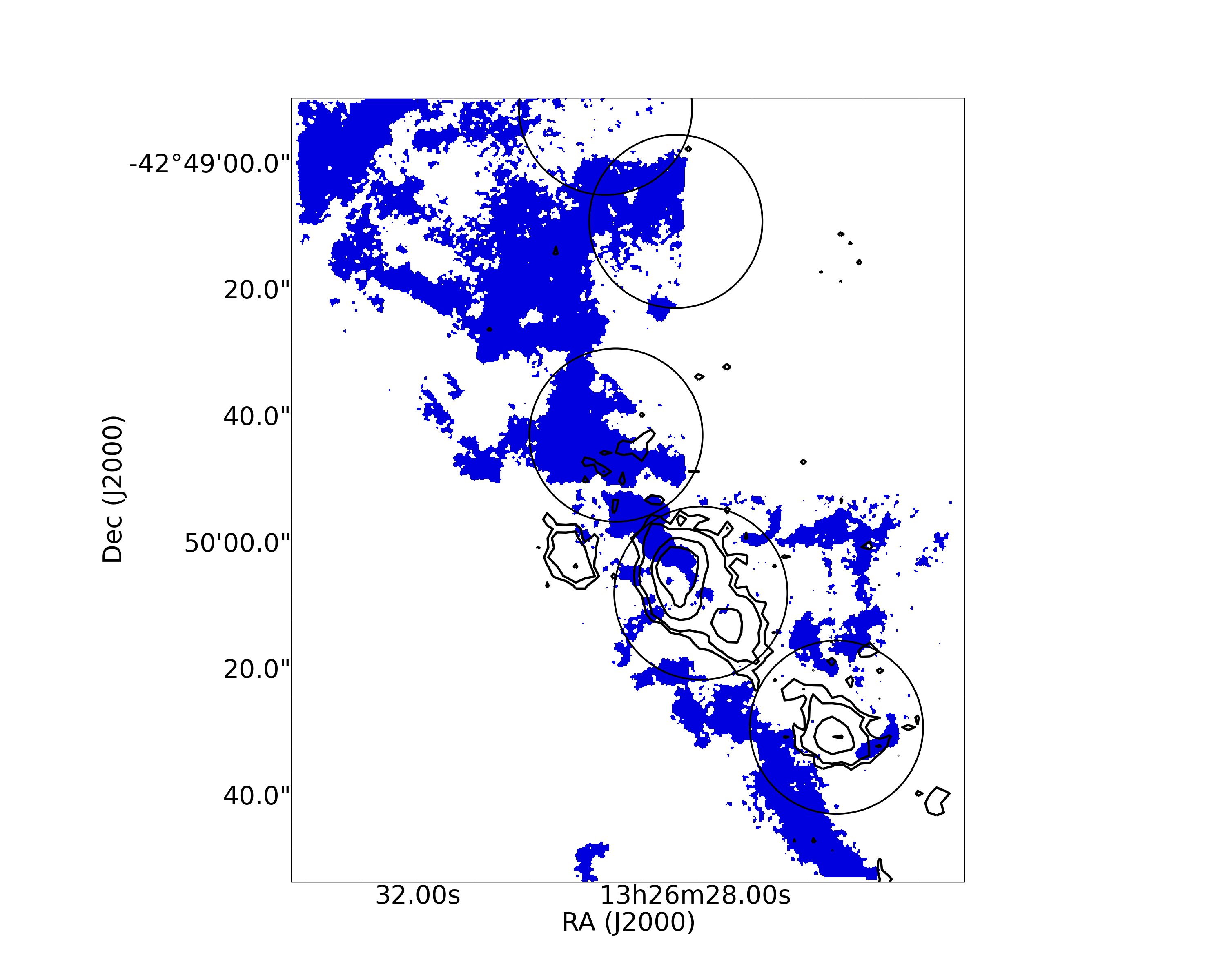}
  \hspace{5mm}
  \includegraphics[height=7cm,trim=510 80 450 170,clip=true]{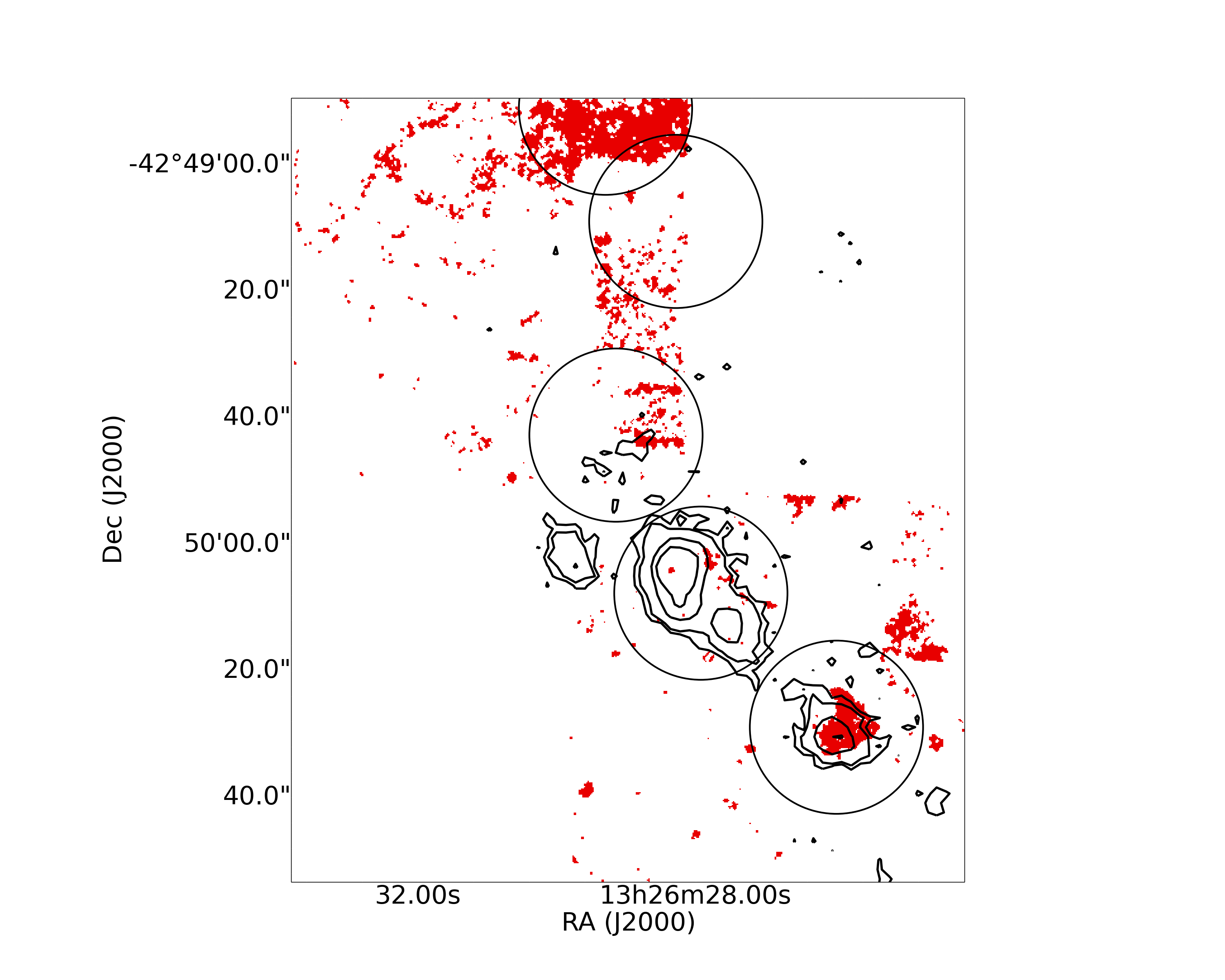}
  \vspace{5mm}
  \includegraphics[height=7cm,trim=160 80 450 170,clip=true]{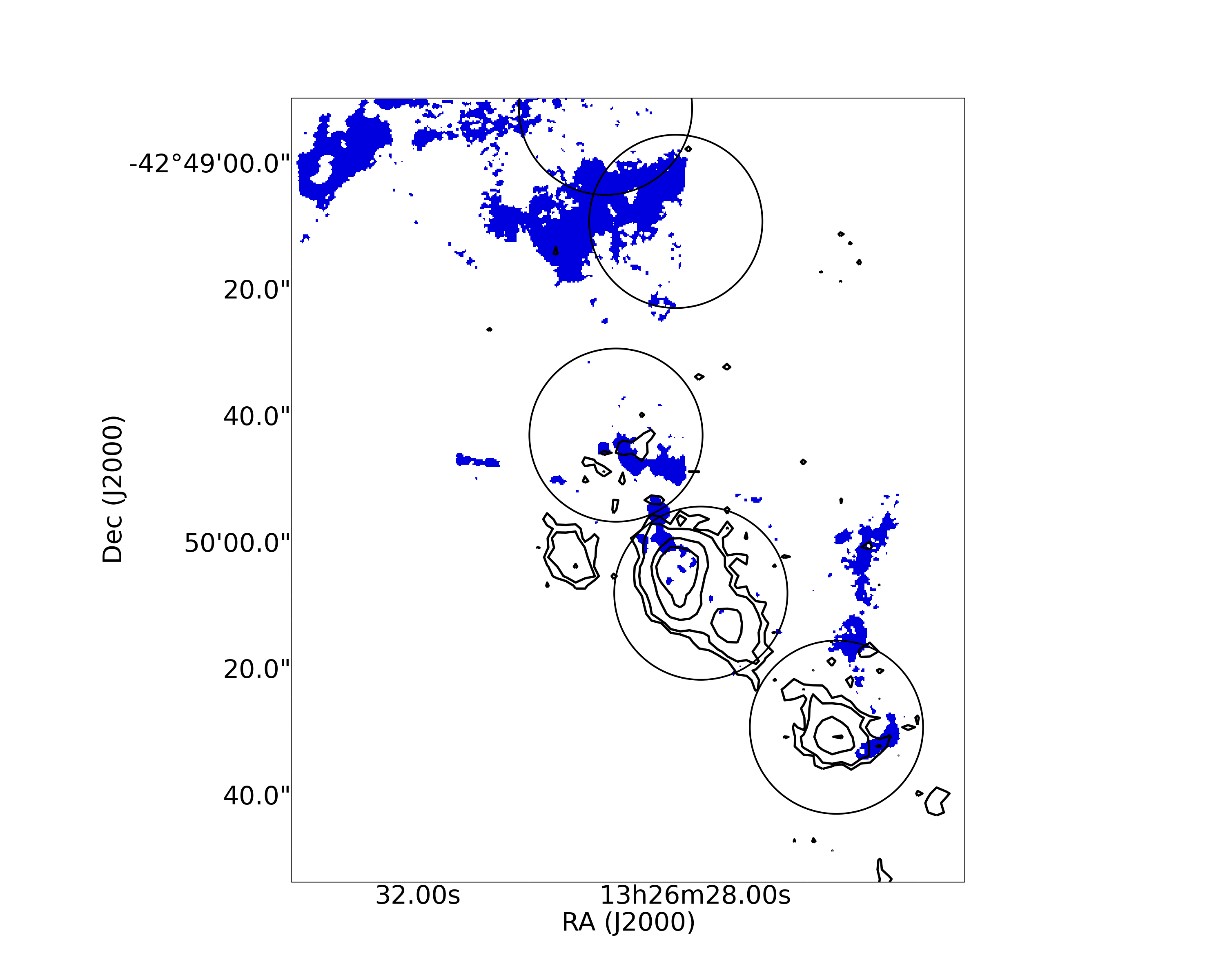}
  \hspace{5mm}
  \includegraphics[height=7cm,trim=510 80 450 170,clip=true]{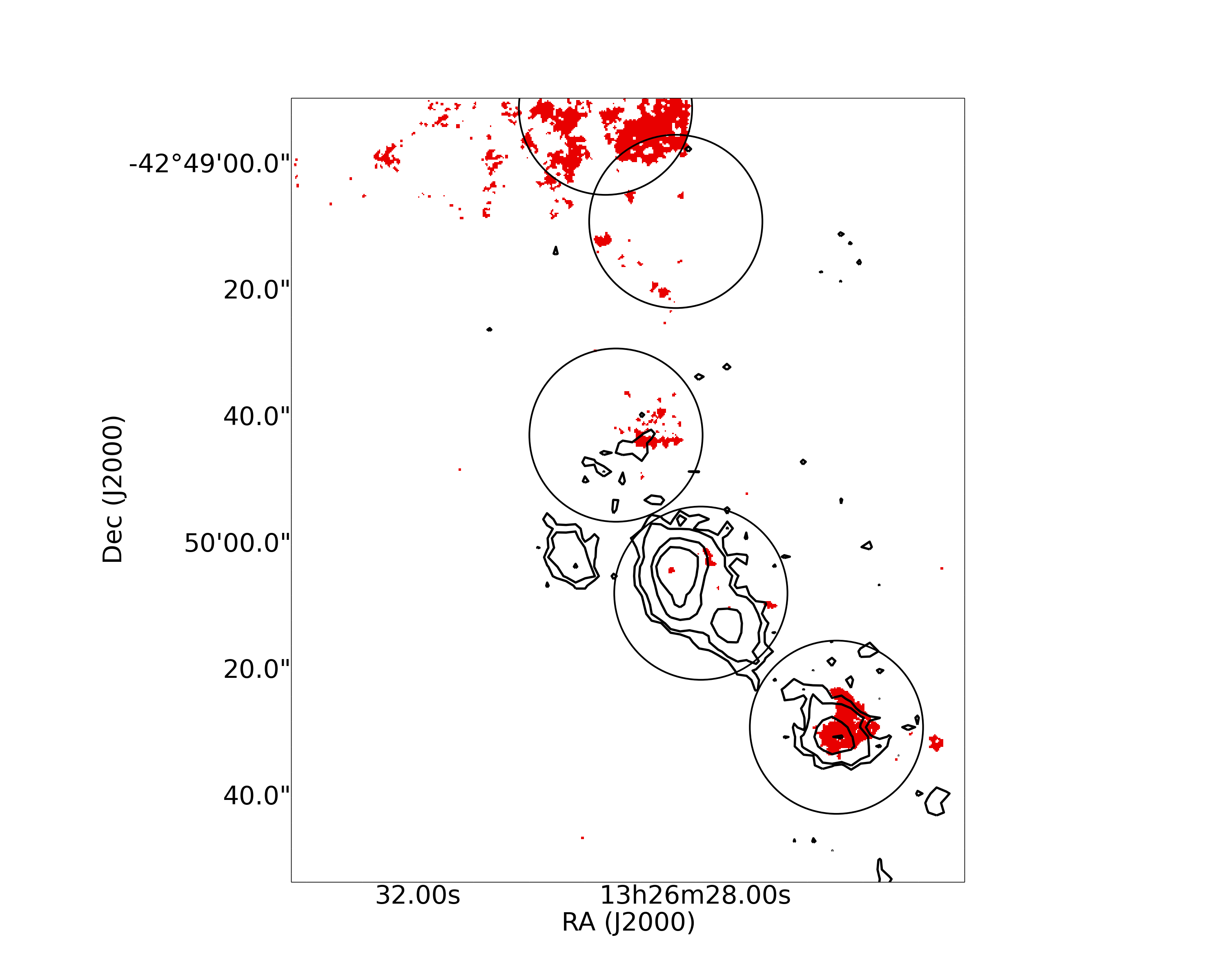}
  \caption{\label{Excitation_MUSE} Map of the excitation processes in the field of view of MUSE. \emph{Top:} RGB map with star formation in green, AGN/shocks in blue, composite in red. The contours show the UV emission from GALEX (black; \emph{left}) and the $H\alpha$-[N\rmnum{2}] emission from MUSE (white; \emph{right}).
\emph{Middle:} Separation of the AGN/shocks excitation (\emph{left}), and the SF+composite excitation (\emph{right}).
\emph{Bottom:} Separation of the AGN/shocks excitation (\emph{left}), and the SF+composite excitation (\emph{right}) for the velocity range of the CO emission ($-256.8<v<-194.5\: km\,s^{-1}$; relative to Centaurus A). The APEX beams are represented by the circles.}
\end{figure*}

   We computed pixel-by-pixel BPT diagrams based on the $H\alpha$, [N\rmnum{2}], $H\beta$, [O\rmnum{3}], [O\rmnum{1}] and [S\rmnum{2}] lines (figure \ref{BPT}; \citealt{Baldwin_1981,Kewley_2006}). We aimed to look at the distribution of the excitation processes. Each line was fitted by a gaussian at each spatial resolution element (see \citealt{Hamer_2014} for the details) in order to measure the flux. Figure \ref{Excitation_MUSE} shows maps of the different regions regarding the excitation process. Most of the filaments seem to be excited by energy injection from the radio jet or shocks. Those large regions contains smaller inclusions that are excited by star formation, similar to that has been claimed by \cite{Santoro_2016}. The star-forming regions spatially coincide with the filaments seen in FUV with GALEX. On the opposite, the regions excited by the radio jet or shocks are brighter in $H\alpha$.
Optical line ratios were then extracted only in the velocity range corresponding to the CO velocities. Those optical diagnostics show that the gas is excited by AGN/shocks in some APEX pointings, and excited by star formation in others. Both processes are at play in position 25. The star-forming region of \cite{Santoro_2016} is spatially and dynamically coincident with molecular gas in position 20.

\section{Discussion}
\label{sec:discussion}

\noindent \textit{The low star formation efficiency} - We found in Figure \ref{KS-law} that the depletion time of the molecular gas in the filaments was longer than the Hubble time. This means a very low efficiency of forming stars with respect to discs. It must be emphasized however that the geometry of the filaments is far from disc-like. Molecular gas is clumpy along filaments that are distributed in a larger volume than discs. Therefore the density of filaments is probably small in the northern shell. The average total gas volumic density must be smaller than the typical volume density in galactic disc. The efficiency is also likely to depend on the pressure and the surface density of star (e.g. \citealt{Blitz_2006}), which is, on average, low in the filaments. With respect to the H\rmnum{1} shell outside the radio jet, the AGN impact has been to convert the gas from the H\rmnum{1} to $H_2$ phase, and to increase its star formation efficiency anyway.

\medskip

\noindent \textit{Metallicity effects} - The CO/H$_2$ conversion factor is known to be linked to the metallicity. At low metallicity, a standard $\alpha_{CO}$ would underestimate the mass of the molecular gas \citep{Wolfire_2010}. In a previous paper \citep{SalomeQ_2016}, we corrected the $\alpha_{CO}$ factor following the method by \cite{Leroy_2013}, by assuming a constant oxygen abundance within the filaments. However, we show in this paper that the physical conditions of the gas are not uniform in the filaments. Therefore the metallicity may be subject to significant variations, leading to a different $\alpha_{CO}$ along the filaments.
Nevertheless, the metallicity ranges deduced from MUSE data are always subsolar \citep{SalomeQ_2016}, so the corresponding $\alpha_{CO}$ should be equal or larger than the standard value we used.
The molecular gas masses derived here are thus at worst underestimated leading to even less efficient star formation in the filaments.

\medskip

\noindent \textit{Star formation tracers} - The star formation rate was estimated by a combination of the IR and FUV emission, the FUV being emitted by massive young OB stars, and the IR being the thermal emission of dust (\citealt{Kennicutt_2012} and references therein). However, dust absorbs both the FUV and $H\alpha$ emission so part of dust heating is likely due to the jet/shock excitation, and not related to star formation. Therefore the star formation rates may be overestimated, leading to the same conclusion as above.

\medskip

\begin{figure}[h!]
  \centering
  \includegraphics[width=\linewidth,trim=25 0 45 25,clip=true]{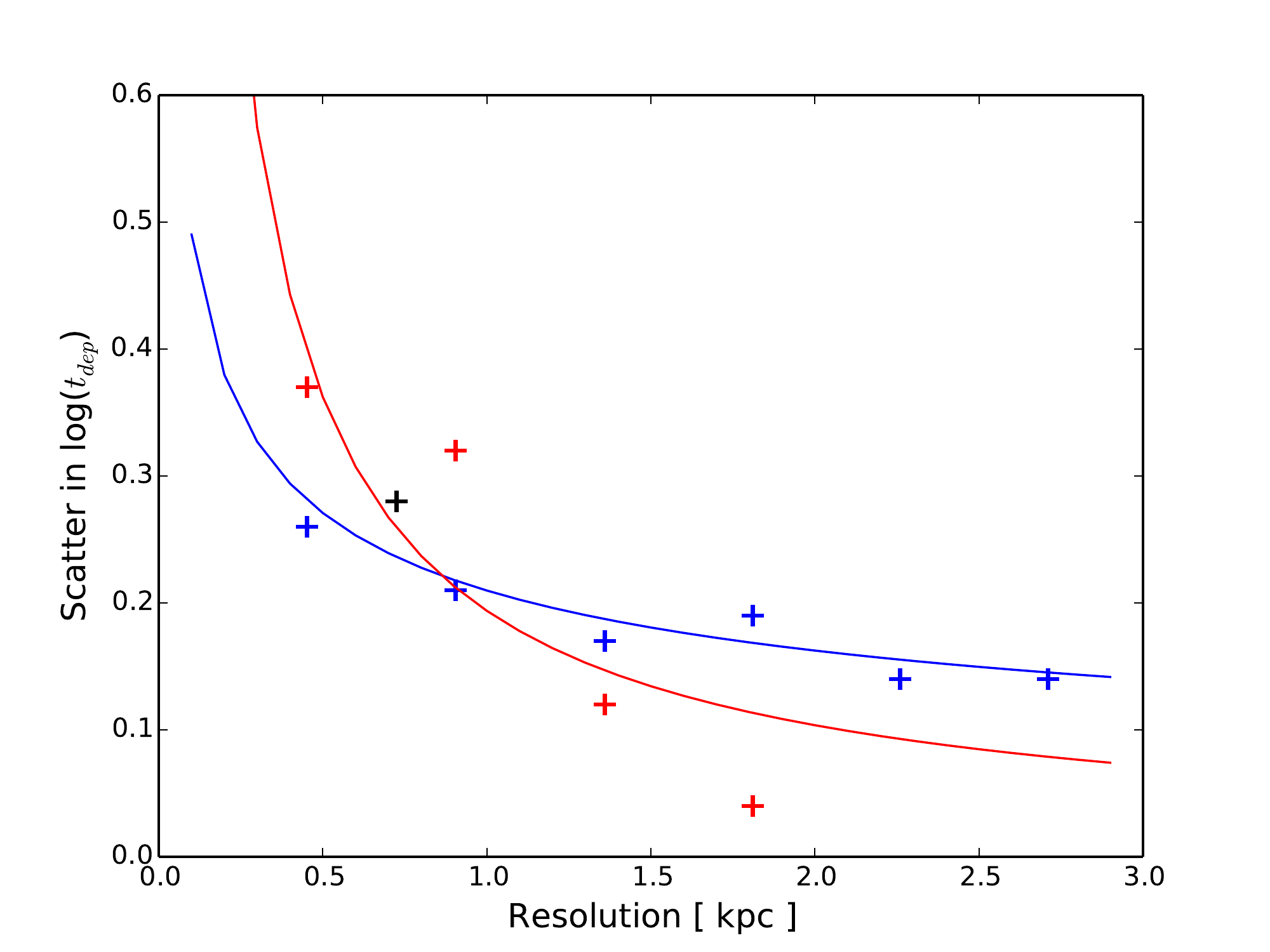}
  \caption{\label{scatter} Rms scatter in $\log t^{mol}_{dep}$ as a function of the resolution in kpc in the filaments. The red and blue lines are the best-fitting model $\sigma(l)=\sigma_{600} \cbra{\frac{l}{600\: pc}}^{-\beta}$ for the eastern ($\beta\sim 0.9$) and western ($\beta\sim 0.35$) part of the filaments \citep{Leroy_2013}.}
\end{figure}

\noindent \textit{Influence of resolution} - Resolution may have effects on the star formation law. \cite{Schruba_2010} showed that, at resolution smaller than a few hundreds parsecs, molecular gas and star formation tracers start to be spatially decorrelated therefore, depending on which regions the aperture is centred (CO emission or star formation tracers), the dependence of $\Sigma_{SFR}$ on $\Sigma_{H_2}$ will vary significantly. By combining APEX beams to simulate lower resolutions, we did not find significant change of the depletion time when changing the resolution, contrary to \cite{Schruba_2010}. \\
We then explored the evolution of the scatter of $\log t_{dep}$ with the resolution. \cite{Leroy_2013} derived the relation $\sigma(l)=\sigma_{600}\cbra{\frac{l}{600\: pc}}^{-\beta}$ where the index $\beta$ is related to the correlation of star formation with the environment. We calculated the scatter of $\log t_{dep}$ in the east and west of the filaments separately, for different combinations of the APEX beams.
We found that the APEX pointings in the eastern filament are forming star rather independently ($\beta\sim 0.9$) whereas star formation in the west seems to be correlated with the environment ($\beta\sim 0.35$) (see Figure \ref{scatter} and \citealt{Leroy_2013} for the details). This tends to indicate that the radio jet could play a more important role in star formation in the western part of the filaments.

\section{Conclusion}

   The Atacama Pathfinder EXperiment (APEX) was used to map the full region of Centaurus A's northern filaments ($5'\times 4'$). This map follows the optical, UV and dusty elongated structure. At a distance of 13.5 kpc from NGC 5128, the observed region is $\sim 4.2\: kpc$-long. The CO(2-1) emission is detected with a resolution of 450 pc in almost all the 31 targeted positions. The undetected regions are lying at the edges of the filaments and deeper observations will determine the accurate limits of emission where the signal is seen to be fading away. The molecular gas mass is found to lie in two separated prominent structures: the eastern region (the brightest) and the western region. Those two structures follow the optically identified east and west arms of the northern filaments. Emission is however found in between these two filaments (region 22) suggesting a possibly larger underlying cold molecular gas reservoir. The southern tip of the filaments, where lie a large UV spot, is much more CO-poor. This region lies at the edge of the H\rmnum{1} shell. The low molecular gas mass may therefore be explained by the gas-poor nature of the region. It might also be explained by a high ionisation of gas, although no clue supports this hypothesis.

   \cite{Charmandaris_2000} pointed and found CO emission in the H\rmnum{1} cloud only (called shell S1 and S2). The CO emission in the north is in fact:
\begin{itemize}
  \item[(i)]   much more extended and following the whole optical filaments
  \item[(ii)]  $\sim 5$ times more massive with $(8.2\pm 0.5)\times 10^7\: M_\odot$ in the whole filaments versus $1.7\times 10^7\: M_\odot$ in Shell S1 only. The filaments molecular content represents about 3-5\% of the total gas mass of the entire galaxy, $1.4-2.7\times 10^9\: M_\odot$ found by Herschel and CO mapping \citep{Parkin_2012}
  \item[(iii)] mostly molecular with a small atomic-to-molecular gas fraction and with the brightest emission being far outside the H\rmnum{1} cloud itself. All the CO(2-1)-detected regions are found to be dust-rich.
\end{itemize}

   Archival Herschel FIR and GALEX UV data were used to determine the star formation rates inside the CO-detected regions. After discussing possible effects of spatial resolution and metallicity, we conclude that the star formation efficiency (SFE) is very low in the northern filaments. Even if traces of recent star formation are claimed in this region (with the presence of young stars), the amount of molecular gas reservoir available is huge compared to standard star formation efficiencies. This suggests that some processes may prevent the star formation to proceed in the cold gas. Statistical studies with higher resolution CO data all along the filaments are underway (ALMA cycle 3 project) and will shed light on possibly local effects. \cite{SalomeQ_2016} analyses of very first ALMA archival data already showed that the molecular gas is clumpy and likely not gravitationally bounded. New ALMA data will be used to study whether this tendency extends to the whole filaments or not, as well as the very first trends determined here in the CO mass distribution function. High spatial resolution mapping of the linewidths distribution will also help identifying the star formation quenching mechanism.

   A possible process that may prevent the molecular gas to form stars is kinetic energy injection from the larger scale dynamics at play in this system. The CO velocity field found with APEX is consistent with the recent findings (\citealt{Oosterloo_2005} and VIMOS data from \citealt{Santoro_2015a}) of shears in the filaments that may be interpreted as a direct effect of the AGN-jet interaction on the underlying gas distribution. The comparison with MUSE data clearly shows the presence of different excitation mechanisms, AGN and/or shocks being the dominant process in the northern filaments, with localised H\rmnum{2} dominated regions. 

  The low star formation efficiency of the molecular filaments, with respect to discs, might also be interpreted by a volumic effect. Discs are much thinner that the gaseous shell in the halo of Centaurus A, and therefore the molecular gas density lower in average, for the same surface density. Maybe the most interesting impact of the radio jet on the gaseous shell is to compress the gas and trigger the phase transition from atomic to molecular gas. Outside of the radio jet, the fraction  of $H_2$ gas was only of the order of 0.2, while the gas has become entirely molecular in the jet. A radio-jet H\rmnum{1} ram pressure stripping scenario is unlikely to explain the unbalanced H$_2$/H\rmnum{1} ratio in the two filaments. In such a case, the molecular gas content is expected to be the same between the western and eastern filaments. Conversely, we found a much larger amount of molecular gas in the eastern filaments (atomic-free). This is certainly the way the AGN and its jets can have a positive feedback effect on the star formation in NGC 5128.

\vspace{5mm}

\noindent \textbf{NOTE ADDED IN PROOF}

\noindent \small{After the submission of this paper, Morganti et al. submitted a similar study with APEX, although with fewer data.}

\vspace{5mm}

\noindent\rule[0.5ex]{\linewidth}{0.5pt}

\begin{acknowledgements}
   We thank Carlos de Breuck, Katharina Immer and APEX operators for their support for the observations. We also thank the referee for helpful comments. \\

   The Atacama Pathfinder EXperiment (APEX) is a collaboration between the Max-Planck-Institut für Radioastronomie, the European Southern Observatory, and the Onsala Space Observatory. \\

   Based on observations made with ESO Telescopes at the La Silla Paranal Observatory under programme 60.A-9341A. \\

   We acknowledge financial support from "Programme National de Cosmologie and Galaxies" (PNCG) of CNRS/INSU, France.
\end{acknowledgements}

\bibliography{Biblio,Biblio_arXiv}
\bibliographystyle{aa}

\clearpage

\clearpage

\appendix
\onecolumn

\section{Additional tables}

\begin{table*}[h]
  \centering
  \footnotesize
  \begin{tabular}{lccccl}
    \hline \hline
    Position & RA (J2000)  & Dec (J2000) & $t_{obs}$ & rms  & Observation run \\
             &             &             &   (min)   & (mK) &                 \\ \hline
    1        & 13:26:16.10 & -42:46:56.0 &   45.2    & 2.15 & 1-2             \\
    2        & 13:26:16.00 & -42:47:17.9 &   17.7    & 3.03 & 1               \\
    3        & 13:26:15.74 & -42:47:39.9 &   17.6    & 2.88 & 1               \\
    4        & 13:26:16.17 & -42:48:01.3 &   14.1    & 3.16 & 1               \\
    5        & 13:26:16.75 & -42:48:21.4 &   17.7    & 2.98 & 1               \\
    6        & 13:26:17.60 & -42:48:42.0 &   17.8    & 2.58 & 1               \\
    7        & 13:26:19.42 & -42:48:52.0 &   14.7    & 3.42 & 1               \\
    8        & 13:26:21.72 & -42:48:56.6 &   14.7    & 3.18 & 1               \\
    9        & 13:26:20.23 & -42:49:12.1 &   14.7    & 2.68 & 1               \\
    10       & 13:26:21.96 & -42:49:22.1 &   17.6    & 2.63 & 1               \\
    11       & 13:26:21.73 & -42:49:44.6 &   17.8    & 2.87 & 1               \\
    12       & 13:26:16.45 & -42:49:55.0 &   53.2    & 1.83 & 1-2             \\
    13       & 13:26:22.05 & -42:50:08.4 &   20.6    & 3.23 & 1               \\
    14       & 13:26:17.77 & -42:50:09.6 &   64.7    & 2.09 & 1-2             \\
    15       & 13:26:21.85 & -42:50:28.6 &   94.4    & 1.81 & 1-2             \\
    16       & 13:26:27.47 & -42:48:50.9 &   17.8    & 2.42 & 1               \\
    17       & 13:26:29.30 & -42:48:51.0 &   14.8    & 3.02 & 1               \\
    18       & 13:26:28.29 & -42:49:08.9 &   17.6    & 3.15 & 1               \\
    19       & 13:26:27.93 & -42:50:07.7 &   87.0    & 1.33 & 1-2             \\
    20       & 13:26:25.98 & -42:50:28.9 &   38.2    & 1.71 & 1-2             \\
    21       & 13:26:24.08 & -42:49:05.4 &   20.7    & 3.01 & 2               \\
    22       & 13:26:24.26 & -42:49:32.6 &   17.7    & 2.56 & 2               \\
    23       & 13:26:26.28 & -42:49:11.8 &    3.5    & 5.05 & 2               \\
    24       & 13:26:26.66 & -42:49:38.1 &    1.5    & 8.58 & 2               \\
    25       & 13:26:29.15 & -42:49:42.7 &   16.8    & 2.53 & 2               \\
    26       & 13:26:13.85 & -42:48:09.6 &   17.7    & 2.59 & 2               \\
    27       & 13:26:13.95 & -42:48:37.1 &   15.5    & 3.14 & 2               \\
    28       & 13:26:23.49 & -42:50:44.5 &   17.7    & 2.51 & 2               \\
    29       & 13:26:22.32 & -42:51:09.8 &   14.2    & 3.12 & 2               \\
    30       & 13:26:21.17 & -42:51:35.9 &    9.7    & 3.42 & 2               \\
    31       & 13:26:20.44 & -42:52:03.6 &    8.9    & 4.39 & 2               \\ \hline
     \end{tabular}
  \caption{\label{table:obs} Journal of observations with APEX. The rms were determined with both polarisations and are given in main beam temperature. The first run was made in September 2015, the second run extends from end of October to December.}
\end{table*}

\begin{table*}[h]
  \centering
  \begin{tabular}{lccccccccc}
    \hline \hline
    Position &  $T_{mb}$ &    $\Delta v$    &    $v_{peak}$    &      $I_{CO}$      &            $L'_{CO}$           &      $M_{H_2}$     \\
             &    (mK)   &  ($km\,s^{-1}$)  &  ($km\,s^{-1}$)  & ($K\,km\,s^{-1}$)  & ($10^5\: K\,km\,s^{-1}\,pc^2$) & ($10^6\: M_\odot$) \\ \hline
    1        &  $<6.45$  &        30        &         -        &      $<0.205$      &            $<0.409$            &       $<0.34$      \\
    2        &   18.61   &  $32.1\pm 5.8$   &  $-198.8\pm 2.3$ &  $0.636\pm 0.094$  &         $1.27\pm 0.19$         &   $1.06\pm 0.16$   \\
    3        &   24.20   &  $32.3\pm 3.8$   &  $-213.0\pm 1.7$ &  $0.831\pm 0.086$  &         $1.66\pm 0.17$         &   $1.39\pm 0.14$   \\
    4        &   40.50   &  $30.3\pm 2.9$   &  $-210.4\pm 1.1$ &  $1.304\pm 0.098$  &         $2.60\pm 0.20$         &   $2.17\pm 0.16$   \\
    5        &   21.89   &  $40.8\pm 5.7$   &  $-208.4\pm 2.3$ &  $0.952\pm 0.106$  &         $1.90\pm 0.21$         &   $1.59\pm 0.18$   \\
    6        &   19.72   &  $24.7\pm 3.8$   &  $-227.6\pm 1.6$ &  $0.518\pm 0.068$  &         $1.03\pm 0.14$         &   $0.86\pm 0.11$   \\
    7        &   22.38   &  $31.1\pm 6.6$   &  $-227.1\pm 2.5$ &  $0.740\pm 0.115$  &         $1.47\pm 0.23$         &   $1.23\pm 0.19$   \\
    8        &   24.29   &  $56.3\pm 5.2$   &  $-205.2\pm 2.4$ &  $1.439\pm 0.123$  &         $2.87\pm 0.25$         &   $2.40\pm 0.21$   \\
    9        &   15.52   &  $52.6\pm 6.2$   &  $-213.9\pm 3.2$ &  $0.869\pm 0.099$  &         $1.73\pm 0.20$         &   $1.45\pm 0.17$   \\
    10       &   56.11   &  $42.9\pm 1.7$   &  $-191.1\pm 0.8$ &  $2.560\pm 0.091$  &         $5.10\pm 0.18$         &   $4.27\pm 0.15$   \\
    11       &   20.92   &  $43.5\pm 5.0$   &  $-191.7\pm 2.3$ &  $0.968\pm 0.100$  &         $1.93\pm 0.20$         &   $1.61\pm 0.17$   \\
    12       &    7.76   &  $47.3\pm 12.4$  &  $-228.3\pm 3.9$ &  $0.391\pm 0.074$  &         $0.78\pm 0.15$         &   $0.65\pm 0.12$   \\
    13       &   13.92   &  $49.8\pm 10.3$  &  $-149.4\pm 4.3$ &  $0.738\pm 0.125$  &         $1.47\pm 0.25$         &   $1.23\pm 0.21$   \\
    14       &   12.74   &  $49.9\pm 12.1$  &  $-204.3\pm 3.1$ &  $0.677\pm 0.104$  &         $1.35\pm 0.21$         &   $1.13\pm 0.17$   \\
    15       &    9.21   & $100.7\pm 14.8$  &  $-203.7\pm 4.5$ &  $0.987\pm 0.096$  &         $1.97\pm 0.19$         &   $1.65\pm 0.16$   \\
    16       &   90.09   &  $79.5\pm 1.4$   &  $-251.5\pm 0.6$ &  $7.625\pm 0.116$  &        $15.20\pm 0.23$         &  $12.71\pm 0.19$   \\
    17       &   88.88   &  $69.5\pm 1.6$   &  $-248.7\pm 0.7$ &  $6.576\pm 0.133$  &        $13.10\pm 0.27$         &  $10.96\pm 0.22$   \\
    18       &   58.00   &  $82.4\pm 2.7$   &  $-256.8\pm 1.2$ &  $5.087\pm 0.151$  &        $10.14\pm 0.30$         &   $8.48\pm 0.25$   \\
    19       &    9.10   &  $46.0\pm 6.4$   &  $-236.3\pm 2.5$ &  $0.445\pm 0.050$  &         $0.89\pm 0.10$         &   $0.74\pm 0.08$   \\
    20       &   22.02   &  $28.9\pm 2.3$   &  $-226.9\pm 1.0$ &  $0.677\pm 0.048$  &         $1.35\pm 0.10$         &   $1.13\pm 0.08$   \\ \hline
    21       &  $<9.04$  &        30        &         -        &      $<0.287$      &            $<0.572$            &       $<0.48$      \\
    22       &    6.34   &  $68.2\pm 17.1$  &  $-257.2\pm 8.8$ &  $0.453\pm 0.109$  &         $0.90\pm 0.22$         &   $0.76\pm 0.18$   \\
    23       &   79.55   &  $62.3\pm 2.8$   &  $-278.4\pm 1.3$ &  $5.274\pm 0.213$  &        $10.51\pm 0.42$         &   $8.79\pm 0.36$   \\
    24       &   69.96   &  $57.7\pm 5.4$   &  $-265.0\pm 2.3$ &  $4.300\pm 0.346$  &         $8.57\pm 0.69$         &   $7.17\pm 0.58$   \\
    25       &   13.99   & $131.1\pm 12.8$  &  $-194.5\pm 5.2$ &  $1.952\pm 0.161$  &         $3.89\pm 0.32$         &   $3.25\pm 0.27$   \\
    26       &   13.62   &  $22.2\pm 4.7$   &  $-210.0\pm 2.4$ &  $0.322\pm 0.063$  &         $0.64\pm 0.13$         &   $0.54\pm 0.11$   \\
    27       &   13.65   &  $23.9\pm 6.9$   &  $-207.7\pm 3.1$ &  $0.347\pm 0.085$  &         $0.69\pm 0.17$         &   $0.58\pm 0.14$   \\
    28       &  $<7.54$  &        30        &         -        &      $<0.240$      &            $<0.478$            &       $<0.40$      \\
    29       &  $<9.37$  &        30        &         -        &      $<0.298$      &            $<0.594$            &       $<0.50$      \\
    30       &   15.93   &  $45.3\pm 8.5$   &  $-295.9\pm 3.6$ &  $0.769\pm 0.124$  &         $1.53\pm 0.25$         &   $1.28\pm 0.21$   \\
    31       & $<13.16$  &        30        &         -        &      $<0.418$      &            $<0.833$            &       $<0.70$      \\ \hline
  \end{tabular}
  \caption{\label{table:specCO} CO luminosities and molecular masses in the regions observed with APEX. The velocities are relative to Centaurus A. $M_{H_2}$ is the molecular mass derived from the CO(2-1) emission with a standard conversion factor and an assumed CO(2-1)/CO(1-0) ratio of 0.55 \citep{Charmandaris_2000}. For non detections, an upper limit at $3\sigma$ has been derived assuming a linewidth of $30\: km.s^{-1}$.}
\end{table*}

\begin{table*}[h]
  \centering
  \begin{tabular}{lcccccccc}
    \hline \hline
    Position &      $M_{H_2}$     &             $SFR$              & $t^{mol}_{dep}$ &    $\Sigma_{H_2}$    &          $\Sigma_{SFR}$          \\
             & ($10^6\: M_\odot$) & ($10^{-5}\: M_\odot\,yr^{-1}$) &      (Gyr)      & ($M_\odot\,pc^{-2}$) &  ($M_\odot\,yr^{-1}\,kpc^{-2}$)  \\ \hline
    1        &      $<0.34$       &              1.96              &     $<17.3$     &       $<2.12$        &       $1.22\times 10^{-4}$       \\
    2        &   $1.06\pm 0.16$   &         $4.86\pm 0.60$         &  $21.8\pm 6.0$  &    $6.60\pm 1.00$    &  $(3.03\pm 0.37)\times 10^{-4}$  \\
    3        &   $1.39\pm 0.14$   &        $11.39\pm 2.35$         &  $12.2\pm 3.7$  &    $8.66\pm 0.87$    &  $(7.10\pm 1.46)\times 10^{-4}$  \\
    4        &   $2.17\pm 0.16$   &        $10.88\pm 0.99$         &  $19.9\pm 3.3$  &   $13.52\pm 1.00$    &  $(6.78\pm 0.62)\times 10^{-4}$  \\
    5        &   $1.59\pm 0.18$   &        $35.99\pm 2.94$         &  $4.42\pm 0.86$ &    $9.90\pm 1.12$    &  $(2.24\pm 0.18)\times 10^{-3}$  \\
    6        &   $0.86\pm 0.11$   &         $5.87\pm 1.58$         &  $14.7\pm 5.8$  &    $5.36\pm 0.69$    &  $(3.66\pm 0.98)\times 10^{-4}$  \\
    7        &   $1.23\pm 0.19$   &         $8.16\pm 1.79$         &  $15.1\pm 5.6$  &    $7.66\pm 1.18$    &  $(5.08\pm 1.12)\times 10^{-4}$  \\
    8        &   $2.40\pm 0.21$   &         $6.95\pm 0.31$         &  $34.5\pm 4.6$  &   $14.95\pm 1.31$    &  $(4.33\pm 0.19)\times 10^{-4}$  \\
    9        &   $1.45\pm 0.17$   &         $4.94\pm 0.20$         &  $29.4\pm 4.6$  &    $9.03\pm 1.06$    &  $(3.08\pm 0.12)\times 10^{-4}$  \\
    10       &   $4.27\pm 0.15$   &         $5.46\pm 0.23$         &  $78.2\pm 6.0$  &   $26.60\pm 0.93$    &  $(3.40\pm 0.14)\times 10^{-4}$  \\
    11       &   $1.61\pm 0.17$   &         $4.94\pm 0.45$         &  $32.6\pm 6.4$  &   $10.03\pm 1.06$    &  $(3.08\pm 0.28)\times 10^{-4}$  \\
    12       &   $0.65\pm 0.12$   &         $4.98\pm 0.68$         &  $13.1\pm 4.2$  &    $4.05\pm 0.75$    &  $(3.10\pm 0.42)\times 10^{-4}$  \\
    13       &   $1.23\pm 0.21$   &              2.69              &  $45.7\pm 7.8$  &    $7.66\pm 1.31$    &       $1.68\times 10^{-4}$       \\
    14       &   $1.13\pm 0.17$   &         $5.22\pm 0.42$         &  $21.6\pm 5.0$  &    $7.04\pm 1.06$    &  $(3.25\pm 0.26)\times 10^{-4}$  \\
    15       &   $1.65\pm 0.16$   &         $4.06\pm 1.89$         &  $40.6\pm 22.9$ &   $10.28\pm 1.00$    &  $(2.53\pm 1.18)\times 10^{-4}$  \\
    16       &  $12.71\pm 0.19$   &        $16.47\pm 2.40$         &  $77.2\pm 12.4$ &   $79.17\pm 1.18$    &  $(1.03\pm 0.15)\times 10^{-3}$  \\
    17       &  $10.96\pm 0.22$   &        $19.98\pm 2.51$         &  $54.9\pm 8.0$  &   $68.27\pm 1.37$    &  $(1.24\pm 0.16)\times 10^{-3}$  \\
    18       &   $8.48\pm 0.25$   &        $21.22\pm 5.05$         &  $40.0\pm 10.7$ &   $52.82\pm 1.56$    &  $(1.32\pm 0.31)\times 10^{-3}$  \\
    19       &   $0.74\pm 0.08$   &              6.24              &  $11.9\pm 1.3$  &    $4.61\pm 0.50$    &       $3.89\times 10^{-4}$       \\
    20       &   $1.13\pm 0.08$   &         $4.76\pm 0.93$         &  $23.7\pm 6.3$  &    $7.04\pm 0.50$    &  $(2.97\pm 0.58)\times 10^{-4}$  \\ \hline
    21       &       $<0.48$      &            $<0.41$             &        -        &       $<2.99$        &      $<2.55\times 10^{-5}$       \\
    22       &   $0.76\pm 0.18$   &            $<0.43$             &    $<176.7$     &    $4.73\pm 1.12$    &      $<2.68\times 10^{-4}$       \\
    23       &   $8.79\pm 0.36$   &         $8.09\pm 1.88$         & $108.7\pm 29.7$ &   $54.76\pm 2.24$    &  $(5.04\pm 1.17)\times 10^{-4}$  \\
    24       &   $7.17\pm 0.58$   &        $13.55\pm 3.97$         &  $52.9\pm 19.8$ &   $44.66\pm 3.61$    &  $(8.44\pm 2.47)\times 10^{-4}$  \\
    25       &   $3.25\pm 0.27$   &         $1.75\pm 0.21$         & $185.7\pm 37.7$ &   $20.25\pm 1.68$    &  $(1.09\pm 0.13)\times 10^{-4}$  \\
    26       &   $0.54\pm 0.11$   &         $5.37\pm 1.02$         &  $10.1\pm 4.0$  &    $3.36\pm 0.69$    &  $(3.35\pm 0.64)\times 10^{-4}$  \\
    27       &   $0.58\pm 0.14$   &         $3.49\pm 0.45$         &  $16.6\pm 6.2$  &    $3.61\pm 0.87$    &  $(2.17\pm 0.28)\times 10^{-4}$  \\
    28       &       $<0.40$      &              1.02              &     $<39.2$     &       $<2.49$        &       $6.35\times 10^{-5}$       \\
    29       &       $<0.50$      &              1.21              &     $<41.3$     &       $<3.11$        &       $7.54\times 10^{-5}$       \\
    30       &   $1.28\pm 0.21$   &         $3.81\pm 0.18$         &  $33.6\pm 7.1$  &    $7.97\pm 1.31$    &  $(2.37\pm 0.11)\times 10^{-4}$  \\
    31       &       $<0.70$      &         $2.44\pm 0.34$         &  $28.7\pm 4.0$  &       $<4.36$        &  $(1.52\pm 0.21)\times 10^{-4}$  \\ \hline

  \end{tabular}
  \caption{\label{table:overview} Molecular gas masses, SFR and $t^{mol}_{dep}$ in regions of $27''$ around the APEX pointings. $\Sigma_{gas}$ and $\Sigma_{SFR}$ are derived assuming the signal is homogeneous in the $27''$ beam.}
\end{table*}

\clearpage

\section{APEX spectra of CO(2-1) emission}

\begin{figure*}[h]
  \centering
  \includegraphics[width=0.275\linewidth,trim=5 20 20 50,clip=true]{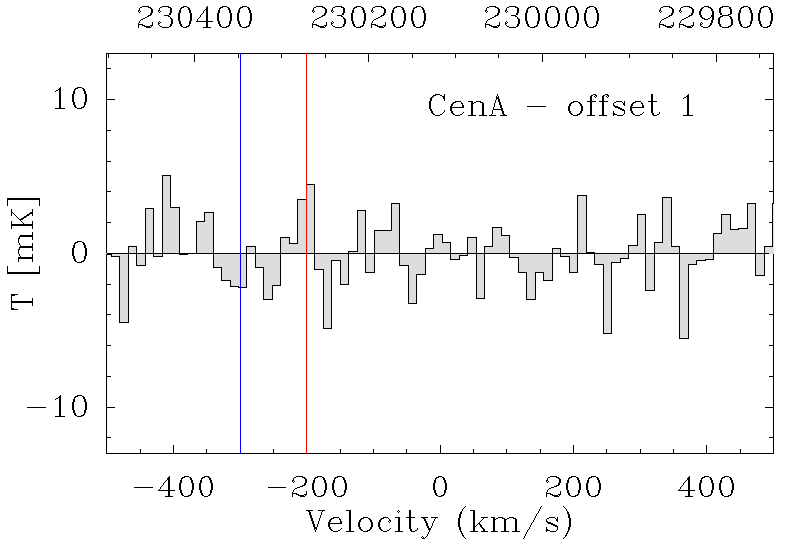}
  \hspace{3mm}
  \includegraphics[width=0.275\linewidth,trim=5 20 20 50,clip=true]{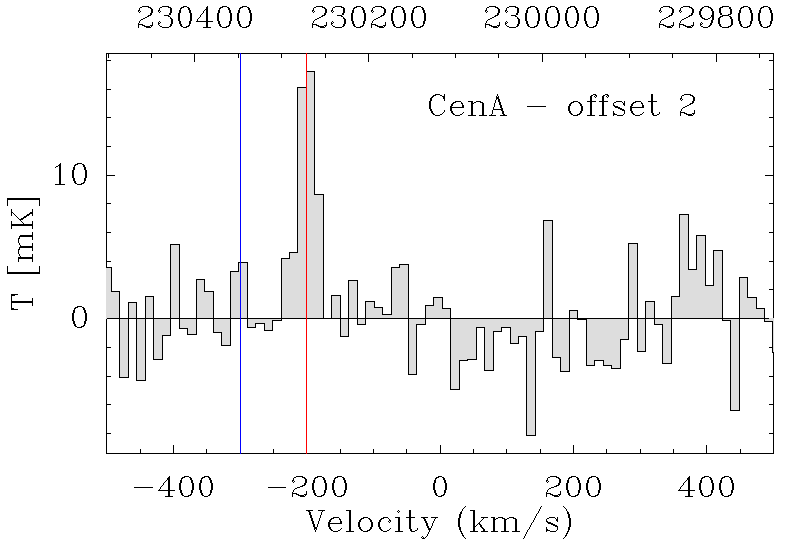}
  \hspace{3mm}
  \includegraphics[width=0.275\linewidth,trim=5 20 20 50,clip=true]{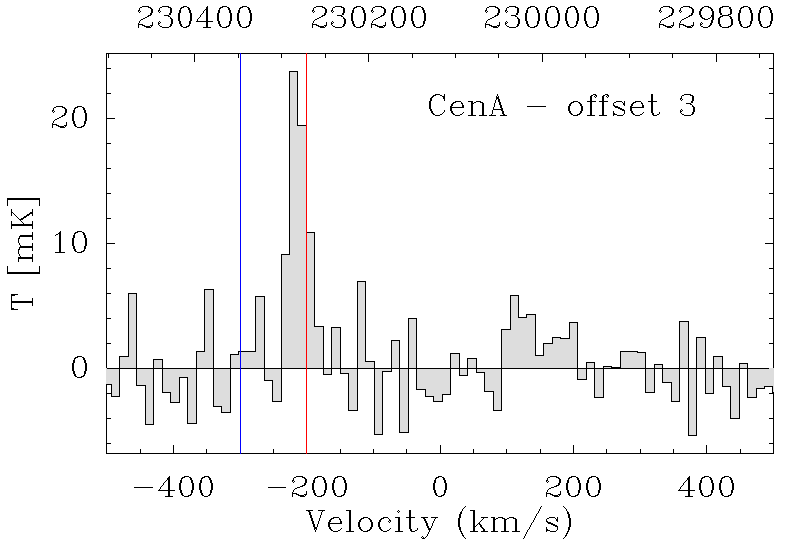}
  \caption{\label{spectra} A sample of the CO(2-1) spectra observed with APEX. The vertical blue and red lines indicate velocities of $-300$ and $-200\: km.s^{-1}$.}
\end{figure*}

\end{document}